\documentclass[preprint, 3p, times]{elsarticle}
\usepackage{graphicx, xcolor}
\usepackage{hyperref}
\usepackage{amsfonts, amsmath, amssymb, mathrsfs, bm, mathtools, stmaryrd}
\usepackage{float, parskip, ulem}
\usepackage[ruled]{algorithm2e}
\usepackage{xparse} 
\DeclarePairedDelimiterX{\Iintv}[1]{\llbracket}{\rrbracket}{\iintvargs{#1}}
\NewDocumentCommand{\iintvargs}{>{\SplitArgument{1}{,}}m}
{\iintvargsaux#1} %
\NewDocumentCommand{\iintvargsaux}{mm} {#1\mkern0.5mu , \mkern0.5mu#2}

\begin{document}

\begin{frontmatter}
\title{High-dimensional reliability-oriented Shapley effects estimation}

\author[1,2,3]{Lucas Monteiro\corref{cor1}}
\ead{lucas.monteiro@onera.fr}
\author[2]{Jérôme Morio}
\ead{jerome.morio@onera.fr}
\author[2]{Julien Demange-Chryst}
\ead{julien.demange-chryst@onera.fr}
\author[4,5]{François Bachoc}
\ead{francois.bachoc@univ-lille.fr}

\cortext[cor1]{Corresponding author}

\affiliation[1]{organization={Institut de Mathématiques de Toulouse, UMR5219 CNRS},
postcode={31062},
city={Toulouse},
country={France}}
\affiliation[2]{organization={ONERA/DTIS, Université de Toulouse},
postcode={F-31055},
city={Toulouse},
country={France}}
\affiliation[3]{organization={ANITI},
city={Toulouse},
country={France}}
\affiliation[4]{organization={Laboratoire Paul Painlevé, UMR8524 CNRS, Université de Lille},
postcode={Cedex 59655},
city={Villeneuve d’Ascq Cedex},
country={France}}
\affiliation[5]{organization={Institut Universitaire de France (IUF)},
country={France}}

\begin{abstract}
This article presents a new estimation scheme for the reliability-oriented Shapley effects when there is a large number of correlated input variables, using a unique sample of failure points. First, we propose a new reformulation of the reliability-oriented closed Sobol indices. This reformulation circumvents the estimation issues associated with existing methods in the literature, but requires estimating densities in potentially high dimensions. Once the reliability-oriented closed Sobol indices have been estimated through these densities, the reliability-oriented Shapley effects are estimated through permutation aggregation to circumvent the issue related to the dimension. In addition, we provide an error estimation procedure relying on the same sample of failing points resulting in a probabilistic error bound, and avoiding any additional call to the model. We illustrate our methodology on numerical use-cases where we compare several density estimators and we propose an application to the monitoring of adaptive importance sampling.
\end{abstract}


\end{frontmatter}

\section{Introduction}\label{sec:intro}

In industrial contexts such as aerospace or the nuclear industry, complex physical systems are studied and reproduced using numerical simulations. These simulations allow to study the behavior of the system in question under different conditions, without having to run the system in practice. For instance, studying an aeronautical system may require flight operations, which are costly and restrictive, and in some cases such as in aerospace, it is even impossible to perform an operation on a satellite, as it is launched only once. From a practical perspective, these simulations are considered as black box models, meaning that the chain of operations performed is so complex that it is preferable to consider the numerical model as an input-output model and to not consider what happens inside it. Furthermore, such numerical codes may be particularly costly in terms of computing time (up to several days CPU) and therefore it is not always possible to request many code evaluations. Moreover, in most industrial contexts, the studied systems are critical, meaning that a malfunction or failure of the system is associated to significant losses, whether material, financial or even human. It is therefore crucial to estimate the probability associated to the failure in order to assess the risk and then to ensure its reliability. In addition, a deeper analysis can be conducted in order to understand the causes of the failure and to prioritize them in terms of importance. In practice, both analyses are achieved through the simulation of extreme configurations, as they provide instances of failure scenarios.

In the context of the uncertainty quantification of a black box model \cite{sullivan2015introduction, ghanem2017handbookuq}, the uncertainties in the inputs are modeled using probability distributions. These uncertainties are then propagated through the model, hence providing an expected value of the output quantity of interest. When one wants to quantify the risk associated to the system, the quantity of interest is the occurrence of the failure, and the reliability analysis consists in estimating the failure probability \cite{morio2015bookrare}. Moreover, to understand the risk and identify its main sources, one may rely of a suited set of methods entitled reliability-oriented sensitivity analysis (ROSA) \cite{chabridon2018these}. These methods consist in estimating some specific indices, such as the well-known Sobol indices \cite{sobol1993sensitivity} for example. However, the latter are limited by the assumption of independence of inputs, which is too strong for most real-world applications. To handle the possible dependence of inputs, the Shapley effects \cite{owen2014shapley, owen2017dependent} have been adapted to the reliability context \cite{ilidrissi2021shap}, and the existing literature provides methods for estimating them in low-dimensional settings (typically with less than 10 inputs) \cite{ilidrissi2021shap, demangechryst2023shaprosa}. However, in practice some models may be higher-dimensional, and thus these methods are no longer adapted. To overcome these limitations, we propose in this article a new method to estimate reliability-oriented Shapley effects for high-dimensional problems. To proceed, we first propose an alternative writing of reliability-oriented closed Sobol indices with densities of potentially large dimension, then we rely on a unique sample of failing points to estimate these densities with appropriate tools. In addition, we provide an error quantification procedure consisting of probabilistic error bounds for these estimates with the same sample of failing inputs. This last contribution is a major benefit for the practitioner as it provides error bounds without requiring any additional call to the model, unlike the more conventional approach consisting in repetitions of the estimation scheme.

The paper is organized as follows. We review in Section \ref{sec:notions} the main notions of reliability analysis and variance-based global sensitivity analysis, and in Section \ref{sec:rosa} we review the main methods of reliability-based sensitivity analysis relying on variance decomposition. Once the gap in the literature has been identified, we develop in Section \ref{sec:targ_shap_nf} the proposed methodology to estimate the target Shapley effects in high-dimensional settings. In addition, we develop in Section \ref{sec:err_quantif} the procedure to estimate the error made on the estimation of the target Shapley effects, taking into account the different sources of error in the proposed estimation scheme, and relying on the same failing sample as for the estimation. Section \ref{sec:num_appli} illustrates the procedure on numerical applications where we compare several density estimators and on an application to adaptive importance sampling where we propose to examine the role of the input variables in exceeding the threshold at each step. Finally, we provide in Section \ref{sec:conclusion} the conclusion and present some perspectives for future developments.

\section{Review on reliability analysis and global sensitivity analysis} \label{sec:notions}

In this section, we first define the notations used throughout the paper, then recall the main concepts of reliability analysis and variance-based global sensitivity analysis to further introduce variance-based reliability-oriented sensitivity indices in Section \ref{sec:rosa}.

\subsection{Set-up and notations}

Throughout this article, we consider a numerical code modeled by a function $\Phi : \mathbb R^d \to \mathbb R$, considered as a black-box, deterministic, costly to evaluate and with no regularity assumptions. The inputs are modeled by the $d$-dimensional random vector $\mathbf X = (X_1, \ldots, X_d)$ with values in $\mathcal X = \prod_{i=1}^d \mathcal X_i \subset \mathbb R^d$ with $\mathcal X_i$ being the support of the random variable $X_i$. Moreover $\mathbf X$ is supposed to be absolutely continuous with probability density function (PDF) $f_{\mathbf X}$ and with no independence assumption. The output of $\Phi$ is written as the random variable $Y = \Phi(\mathbf X)$ with values in $\mathbb R$ and is only supposed to be square integrable.

We denote by $[d]$ the set $\{1, \ldots, d\}$ and by $\mathcal P(d)$ its power set. For any $u \in \mathcal P(d)$, we write $|u|$ for its cardinality, $-u = [d] \backslash u$ for its complementary and $\mathbf X_u = (X_i)_{ i \in u}$ for the associated input vector with values in $\mathcal X_u = \prod_{i \in u} \mathcal X_i$. For a random variable $Z$ with density $f_Z$, we denote its expectation and variance operators by $\mathbb E_{f_Z}$ and $\mathbb V_{f_Z}$ respectively, or by $\mathbb E$ and $\mathbb V$ if there is no ambiguity.

\subsection{Reliability analysis} \label{sec:reliability_analysis}

As described in the introduction, the reliability analysis of a black-box numerical model consists in estimating the failure probability. Without loss of generality, the failure event is defined by $\{\Phi(\mathbf X) > t\}$ with $t \in \mathbb R$ and its associated probability is

\begin{equation}
    p_t := \mathbb P(\Phi(\mathbf X) > t) = \int_{\mathcal X} \mathbf 1_{\Phi(x) > t} f_{\mathbf X}(x) dx = \mathbb E[\mathbf 1_{F_t} (\mathbf X)],
\end{equation}

where $F_t = \{x \in \mathcal X \; | \; \Phi(x) > t\}$ is the failure domain.

Estimating this probability is challenging because the failure is generally a rare event, meaning that $p_t \ll 1$. Since $\Phi$ is costly to evaluate, crude Monte Carlo (MC) sampling is not adapted as it requires too many evaluations of the model $\Phi$ to get a satisfying precision. For instance, if $p_t \approx 10^{-r}$,  then $10^{r+2}$ samples are required to get an error of $10 \%$. To overcome this difficulty, more suited methods can be considered such as importance sampling \cite{bucklew2010raresimulation}, subset sampling \cite{au2001subsetsampling}, moving particles \cite{proppe2021local}, FORM/SORM \cite{hasofer1974form, breitung1984asymptoticapprox} or surrogate-based procedures \cite{echard2011akmcs, echard2013iskriging, huang2016akss}. In particular, many methods rely on failing samples, meaning samples $\widetilde{\mathbf X}^{(n)}$ of $\mathbf X$ satisfying $\Phi(\widetilde{\mathbf X}^{(n)}) > t$ and distributed according to the conditional density $f_{\mathbf X | \mathbf X \in  F_t}$, denoted $f_{\mathbf X | F_t}$, and defined by

\begin{equation}f_{\mathbf X | F_t}(x) := \frac{f_{\mathbf X}(x) \mathbf 1_{F_t}(x)}{p_t}, \qquad \forall x \in \mathcal X.
\end{equation}

Indeed, their mechanism is such that they will naturally generate and use some samples $\widetilde{\mathbf X}^{(n)}$ to estimate $p_t$. For instance, importance sampling (IS) can be combined with Markov Chain Monte Carlo to generate samples from an appropriate auxiliary distribution \cite{au2004ismcmc}. Another example is adaptive sampling which is based on sequential Monte Carlo to gradually sample closer and closer to the failure domain $F_t$ \cite{cerou2007adaptive, cerou2012sequentialmc, au2001subsetsampling}. Overall, from these estimation procedures for $p_t$, it is possible to recover a failing $N$-sample $(\widetilde{\mathbf X}^{(n)})_{n=1}^N \sim f_{\mathbf X | F_t}$.

\subsection{Variance-based global sensitivity analysis} \label{sec:var_gsa}

Global sensitivity analysis (GSA) is concerned by the problem of allocating the uncertainty of the output into the different sources of uncertainty in the inputs \cite{saltelli2004sensitivity, saltelli2008global}. A such analysis can be done for several purposes, such as input hierarchisation or dimensionality reduction.

In case where the inputs are independent, the functional Hoeffding decomposition \cite{hoeffding1948class, sobol1969multiquadrature} is unique and leads to the functional ANOVA decomposition of $\mathbb V(Y)$. From this decomposition, we can define the Sobol indices for variance-based GSA \cite{sobol1993sensitivity}. In particular, the $i^{th}$ first-order index quantifies the portion of $\mathbb V(Y)$ explained only by $X_i$, and the $i^{th}$ total-order index quantifies the portion of $\mathbb V(Y)$ explained by $X_i$ in interaction with all other variables. The Sobol indices are popular sensitivity indices as they are convenient to interpret and there exist many procedures to estimate them \cite{demangechryst2024these, daveiga2024efficientsobol, boucharif2026sobol}. When the inputs are dependent, the Hoeffding decomposition mentioned above is no longer unique, which implies that the Sobol index associated with a subgroup of variables $u \in \mathcal P(d)$ can no longer be interpreted as the percentage of variance explained by $\mathbf X_u$. When the inputs are dependent, Sobol indices can still be used and estimated \cite{saltelli2001modelfree, ratto2001importance, kucherenko2012global, xu2007global, daveiga2009local}, but they are still not interpretable as such. Other approaches are possible, such as those based on block independence \cite{jacques2006corr, gilquin2015replication, broto2019indep}, ANCOVA decomposition  \cite{li2010global, sudret2013analysis} or isoprobabilistic transformations \cite{mara2012variance, mara2015nonparam, tarantola2017variancedep}. Finally, it has been proposed to extend the estimation of total effects, taking into account dependent and constrained inputs \cite{borgonovo2024total}.

Another type of indices in variance-based sensitivity analysis is the Shapley effects, which originally from game theory \cite{shapley1953value, roth2005shapley}. Considering the set of players $[d]$ of a game, a cost function $c : \mathcal P(d) \to \mathbb R$ provides the worth $c(u)$ of a group of players $u \in \mathcal{P}(d)$ according to its contribution. The Shapley values represent the unique allocation of individual contribution to the $d$ players fulfilling the four necessary axioms defined by \cite{shapley1953value} to obtain a fair allocation of the game's gains. Shapley values have been adapted to the context of GSA \cite{owen2014shapley, song2016shapley, owen2017dependent} under the name "Shapley effects" by considering the input variables $X_1, \ldots, X_d$ as the players and by setting $c(u) = \text{S}_u^{\text{c}}$, where $\text{S}_u^{\text{c}}$ is the closed Sobol index associated to the group $u \in \mathcal P(d)$ and defined by

\begin{equation}\label{eq:closed_sobol}
  \text{S}_u^{\text{c}} = \frac{\mathbb V (\mathbb E [\Phi(\mathbf X) |  \mathbf X_u])}{\mathbb V(\Phi(\mathbf X))}.  
\end{equation}

Using $\text{S}_u^{\text{c}}$ defines a valid cost function (see \cite{owen2014shapley, song2016shapley, owen2017dependent} for a discussion of the required properties) and the associated Shapley effect $\text{Sh}_i$ associated to the variable $X_i$ is defined by

\begin{equation}\label{eq:shap_subset}
    \text{Sh}_i = \frac{1}{d} \sum_{u \in [d] \backslash \{i\}} \binom{d-1}{|u|}^{-1}  (\text{S}_{u \, \cup \, \{i\}}^{\text{c}} - \text{S}_u^{\text{c}} ).
\end{equation}

The quantity $\text{S}_{u \cup \{i\}}^{\text{c}} - \text{S}_u^{\text{c}}$ measures the marginal variability of $Y$ explained by the addition of the variable $X_i$ to the set of variables $\mathbf X_u$. The Shapley effect can deal with correlated input variables and take into account both dependence and interaction effects between the variables, and they can be interpreted as the average percentage of variance explained by the variable $X_i$, taking into account all possible contributions of $X_i$ to the different groups $u \in \mathcal P(d)$. The Shapley effects are convenient to interpret as percentages since they are positive and sum to one Nevertheless, we note that they present some limitations. First, they cannot discriminate between pure interaction and dependence effects \cite{chastaing2015sensitivity, ilidrissi2025hoeffdingdecomp, ferrere2025multivariatebernoulli}. Second, they may affect a positive value to an exogenous variable, whereas its effect on the output is null. This last limitation has recently motivated the development of new sensitivity indices \cite{herin2024pme, foucault2023pmetomo}.

The Shapley effect $\text{Sh}_i$ can also be rewritten as the average marginal contribution of the variable $X_i$ over all the possible different orders at which $X_i$ joins a group of variables \cite{castro2009polynomialshapley, song2016shapley},

\begin{equation}\label{eq:shap_perm}
   \text{Sh}_i = \frac{1}{d!} \sum_{\pi \in \mathfrak S_d} \text{S}_{P_i(\pi) \, \cup \, \{i\}}^{\text{c}} - \text{S}_{P_i(\pi)}^{\text{c}} , 
\end{equation}

where $\mathfrak S_d$ is the set of permutations of $[d]$ representing the different possible orders in which the marginal contributions are computed, and $P_i(\pi)=\{\pi(j) \; | \; j \in \Iintv{1, \pi^{-1}(i) - 1}\}$ is the set of indices strictly before $i$ in permutation $\pi$. For instance, for $d=3$, consider the two permutations $\pi = (1, 3, 2)$ and $\pi' = (2, 3, 1)$. For $\pi$, we obtain $\text{S}_{P_1(\pi) \, \cup \, \{1\}}^{\text{c}} - \text{S}_{P_1(\pi)}^{\text{c}} = \text{S}_{\{1\}}^{\text{c}} - \text{S}_{\emptyset}^{\text{c}}$, which is the marginal contribution of $X_1$ alone, while for $\pi'$ the quantity $\text{S}_{P_1(\pi') \, \cup \, \{1\}}^{\text{c}} - \text{S}_{P_1(\pi')}^{\text{c}} = \text{S}_{\{2, 3, 1\}}^{\text{c}} - \text{S}_{\{2, 3\}}^{\text{c}}$ corresponds to the added contribution of $X_1$ to the remaining variables $X_2, X_3$.

Estimation procedures for Shapley effects have been discussed in depth \cite{iooss2019shapley, goda2021shapley, broto2020shapley, benoumechiara2019shapley}. The most classical ones may be summarized in two steps: first, the indices $\text{S}_u^{\text{c}}$ are estimated by double Monte Carlo or by Pick-Freeze \cite{demangechryst2024these}, and second, the marginal contributions $\text{S}_{u \cup \{i\}}^{\text{c}} - \text{S}_u^{\text{c}}$ are aggregated either by computing $\text{S}_u^{\text{c}}$ for all subsets $u$ of $[d]$ as in \eqref{eq:shap_subset}, or by Monte Carlo sampling on permutations to estimate the expectation in \eqref{eq:shap_perm} \cite{castro2009polynomialshapley, song2016shapley}. When the dimension increases, estimating the indices $\text{S}_u^{\text{c}}$ at any order becomes cumbersome as their number increases exponentially with $d$, and due to their combinatorial nature, Shapley effects require a too large number of closed Sobol' indices. To overcome these difficulties, some recent works have proposed to use the Mobius inverse \cite{plischke2021shapley}, or to extend their estimation to Gaussian linear cases specifically \cite{broto2021gl, broto2022block} when the covariance matrix is block-diagonal. From the field of explainable artificial intelligence, it has also been proposed to sample permutations using Quasi Monte Carlo \cite{mitchell2022perm}, to rewrite the Shapley problem in matrix form \cite{lundberg2017shap, musco2025shap, olsen2026shap} using random linear algebra \cite{martinsson2021rndmlinalgebra}, to use latent manifold representation \cite{hu2024manifold} or to use projected random forests to approximate \eqref{eq:shap_subset} through a more efficient sampling of subsets $u \in \mathcal P(d)$ \cite{benard2022shaff}.

\section{Reliability-oriented Sensitivity analysis} \label{sec:rosa}

When studying the failure of a system, estimating the failure probability allows to quantify its risk, but it brings no information on why the failure occurs and how the uncertainty in the input variables is related to the failure. In this context, performing a GSA on the quantity $Y$ may prove to be unadapted and hence uninformative. This has motivated reliability-oriented sensitivity analysis (ROSA) \cite{chabridon2018these} in order to study the influence of the uncertainty in input variables on the failure event. For such a study, two methodologies have been proposed: conditional sensitivity analysis (CSA) and target sensitivity analysis (TSA) \cite{marrel2021target, raguet2018targetconditionalsensitivityanalysis}. The former studies the influence of the input variables only within the failure domain, while the latter studies the influence of the input variables over the occurrence of the failure on the whole input domain. In this section, we present the variance-based TSA methods.  Note that, although out of our scope, other methods and indices exist, in particular those based on FORM/SORM \cite{papaioannou2021form, papaioannou2025formgsa}, dependence measures such as HSIC or Borgonovo indices \cite{marrel2021target, raguet2018targetconditionalsensitivityanalysis, daveiga2015gsa, borgonovo2007importance, derennes2019these} and SHAP-based approaches \cite{pramudita2024rosa}.

\subsection{Variance-based TSA for independent inputs}

Following the presentation of Section \ref{sec:var_gsa}, variance-based ROSA relies on Sobol indices by considering the quantity of interest $\mathbf 1\{\Phi(\mathbf X) > t\} = \mathbf 1_{F_t}(\mathbf X)$ instead of $\Phi(\mathbf X)$. The $i^{th}$ first-order target Sobol index $\text{T-S}_i$ for $X_i$ \cite{wei2012efficientsampling} is defined by

\begin{equation}\label{eq:target_first_sobol_init}
    \text{T-S}_i  = \frac{\mathbb V (\mathbb E[\mathbf 1_{F_t}(\mathbf X) | X_i])}{\mathbb V(\mathbf 1_{F_t}(\mathbf X))},
\end{equation}

and the $i^{th}$ total-order target Sobol index $\text{T-S}_{-i}$ for $X_i$ is defined by

\begin{equation}\label{eq:target_total_sobol_init}
    \text{T-S}_{-i}  = 1 - \frac{\mathbb V (\mathbb E[\mathbf 1_{F_t}(\mathbf X) | \mathbf X_{-i}])}{\mathbb V(\mathbf 1_{F_t}(\mathbf X))}.
\end{equation}

In both \eqref{eq:target_first_sobol_init} and \eqref{eq:target_total_sobol_init}, the denominator is $\mathbb V(\mathbf 1_{F_t}(\mathbf X)) = p_t (1 - p_t)$ since $\mathbf 1_{F_t}(\mathbf X)$ is a Bernoulli random variable with probability $p_t$, corresponding to the occurrence of the failure event. In case where inputs are independent, $\text{T-S}_{i}$ and $\text{T-S}_{-i}$ are interpretable sensitivity indices, as described earlier. To estimate these quantities, it has been at first proposed to combine Monte Carlo with importance sampling in case where the failure probability is small \cite{wei2012efficientsampling}. Other procedures rely on an alternative writing of target first and total order Sobol indices \cite{xiao2017structural}, given respectively by

\begin{equation}\label{eq:target_first_sobol}
    \text{T-S}_i = \frac{p_t}{1 - p_t} \mathbb V_{f_{X_i}}\left( \frac{f_{X_i | F_t}(X_i)}{f_{X_i}(X_i)} \right),
\end{equation}

and

\begin{equation}\label{eq:target_total_sobol}
    \text{T-S}_{-i} = 1 - \frac{p_t}{1 - p_t} \mathbb V_{f_{\mathbf X_{-i}}}\left( \frac{f_{\mathbf X_{-i} | F_t}(\mathbf X_{-i})}{f_{\mathbf X_{-i}}(\mathbf X_{-i})} \right),
\end{equation}

where $f_{X_{i} | F_t}$ is the marginal PDF of $f_{\mathbf X | F_t}$ with respect to $X_i$ while $f_{\mathbf X_{-i} | F_t}$ is the marginal PDF of $f_{\mathbf X | F_t}$ with respect to all variables except $X_i$. The integral representations of \eqref{eq:target_first_sobol} and \eqref{eq:target_total_sobol} are provided in \ref{appendix:integralexp}.

To estimate the target Sobol indices $\text{T-S}_i$ and $\text{T-S}_{-i}$, the main difficulty now shifts towards estimating the probability density functions $f_{X_i | F_t}$ and $f_{\mathbf X_{-i} | F_t}$. The proposed procedures \cite{yun2018rosa, perrin2019rosa} first and foremost rely on failing samples from the reliability analysis, obtained with MC sampling, subset sampling or moving particles, and then approximate the conditional densities using either Edgeworth expansions \cite{yun2018rosa} or non-parametric methods \cite{perrin2019rosa}.

\subsection{Variance-based TSA for dependent inputs}

When input variables are correlated, the target Sobol indices lose their interpretability power. To overcome this limitation in the variance-based paradigm, it has been recently proposed \cite{ehre2024rosa} to extend the works of \cite{mara2012variance, mara2015nonparam} for ROSA by relying on an adapted sampling procedure to obtain failing samples and then applying hierarchical isoprobabilistic transformations to compute the conditional densities $f_{X_i | F_t}$ and $f_{\mathbf X_{-i} | F_t}$. This approach allows to recover interaction effects along with pure dependence effects, which lead to several indices for each input variable to circumvent the issue of interpretability.

Another solution is to use the Shapley effects as defined in \eqref{eq:shap_subset} with an alternative cost function suited for TSA \cite{ilidrissi2021shap}. Indeed, one can consider as cost function the closed Sobol indices defined in \eqref{eq:closed_sobol}, but with $\mathbf 1_{F_t}(\mathbf X)$ as the quantity of interest instead of $\Phi(\mathbf X)$, thus giving the closed target Sobol indices $\text{T-S}^{\text{c}}_u$, 

\begin{equation}\label{eq:closed_target_sobol}
  \forall u \in \mathcal P(d), \qquad \text{T-S}_u^{\text{c}} = \frac{\mathbb V (\mathbb E [\mathbf 1_{F_t}(\mathbf X) |  \mathbf X_u])}{\mathbb V (\mathbf 1_{F_t}(\mathbf X))}.
\end{equation}

Again, using $\text{T-S}_u^{\text{c}}$ defines a valid cost function, as shown in \cite{ilidrissi2021shap} for the broader family of distance-based Shapley effects, where the $\ell_2$ distance allows to recover the closed target Sobol indices $\text{T-S}_u^{\text{c}}$. One should note that the quantity $\text{T-S}_u^{\text{c}}$ is also related to quantile-oriented sensitivity analysis methods \cite{kucherenko2019quantile, fort2016contrast} such as the CRPS indices \cite{borgonovo2021crps} or rank-based estimators \cite{gamboa2022rank}. In the following we denote $\text{T-Sh}_i$ the target Shapley effect associated to $X_i$, using the cost function $c(u) = \text{T-S}_u^{\text{c}}$. The proposed estimation schemes in \cite{ilidrissi2021shap} are based on MC sampling with nearest-neighbor approximations, following the methodology developed in \cite{broto2020shapley} for the GSA Shapley effects. However, these estimations schemes are inappropriate when $p_t \ll 1$ as they require a too large number of calls to the model $\Phi$.

To remedy these limitations, a recent methodology \cite{demangechryst2023shaprosa} introduced a set of new estimation schemes based on importance sampling (IS). After having estimated the failure probability with an IS procedure, the resulting sample can be reused to estimate the target Shapley effects without any additional calls to $\Phi$. It has been shown that when the auxiliary distribution is well-suited to the problem, the IS-based approach provides lower-variance estimators than crude MC when $p_t \ll 1$. However, this method suffers from the curse of dimensionality as it relies on a nearest-neighbor approximation. As a consequence, we propose in this article a new estimation scheme for the target Shapley effects, together with an estimation procedure for the estimation error, when the dimension exceeds 10, and when only a failing sample $(\widetilde{\mathbf X}^{(n)})$ is available.

Our contributions are:

\begin{itemize}
    \item A new rewriting of the closed target Sobol index $\text{T-S}_u^{\text{c}}$ for any $u \in \mathcal P(d)$.

    \item An estimation scheme for $\text{T-Sh}_i$ through density estimation, only with the failing $N$-sample $(\widetilde{\mathbf X}^{(n)})$ (see Section \ref{sec:reliability_analysis}).

    \item A procedure to estimate the error's estimation of $\text{T-Sh}_i$, again only with the same $N$-sample $(\widetilde{\mathbf X}^{(n)})$.
\end{itemize}

\section{Target Shapley effects estimation through density estimation} \label{sec:targ_shap_nf}

In this section, we present a new procedure to estimate the target Shapley effects for moderate-to-large dimensional problems. In this context, we assume that a prior reliability analysis has already been done, thus providing an estimator $\widehat{p}_t$ of $p_t$ and a failing $N$-sample $\widetilde{\mathbf X}^{(n)} \sim f_{\mathbf X | F_t}$ (see Section \ref{sec:reliability_analysis}). The proposed methodology is the following. First, we propose a new rewriting of the closed target Sobol indices \eqref{eq:closed_target_sobol} for any $u \in \mathcal P(d)$, following the same principle as in \eqref{eq:target_first_sobol} and \eqref{eq:target_total_sobol}. Then we propose a estimation scheme for $\text{T-S}_u^{\text{c}}$ relying on the available failing $N$-sample $\widetilde{\mathbf X}^{(n)}$. Finally, we rely on a well-suited aggregation procedure to avoid the computational burden due to the dimension. As a result, this methodology allows to obtain estimates of the target Shapley effects without additional calls to $\Phi$ than those used to estimate $p_t$.

\subsection{Rewriting of closed target Sobol indices}

We propose a new approach for estimating reliability-based Shapley indices based on a reformulation of $\text{T-S}_u^{\text{c}}$. Indeed, in current estimation schemes, the main limitation lies in the use of nearest-neighbors approximation to estimate $\mathbb V (\mathbb E [\mathbf 1_{F_t}(\mathbf X) |  \mathbf X_u])$ in \eqref{eq:closed_target_sobol}. To circumvent this issue, we instead propose to write $\text{T-S}_u^{\text{c}}$ as

\begin{equation} \label{eq:rewriting_var}
    \forall u \in \mathcal P(d), \qquad \text{T-S}_u^{\text{c}} = \frac{p_t}{1-p_t}  \mathbb V_{f_{\mathbf X_u}} \left(\frac{f_{\mathbf X_u | F_t}(\mathbf X_u)}{f_{\mathbf X_u}(\mathbf X_u)} \right).
\end{equation}

In what precedes, $f_{\mathbf X_u | F_t}$ is the marginal PDF of $f_{\mathbf X | F_t}$ with respect to $\mathbf X_u$ and $f_{\mathbf X_u}$ is the marginal PDF of $f_{\mathbf X}$ with respect to $\mathbf X_u$. The proof of \eqref{eq:rewriting_var} is in \ref{appendix:proofs}.

Introducing the density $f_{\mathbf X_u | F_t}$ in \eqref{eq:rewriting_var} allows to circumvent the required nearest-neighbor approximation. However, relying directly on \eqref{eq:rewriting_var} for the estimation is not appropriate in our context since a sample $\mathbf X_u^{(n)} \sim f_{\mathbf X_u}$ is required to estimate $\mathbb V_{f_{\mathbf X_u}}$ while we only have access to a sample $\widetilde{\mathbf X}_u^{(n)} \sim f_{\mathbf X_u | F_t}$. Furthermore, using this last sample would not allow for a precise and numerically stable estimation of the density $f_{\mathbf X_u | F_t}$. As a consequence, we propose in \eqref{eq:rewriting_exp} to rewrite this variance as an expectation with respect to $f_{\mathbf X_u | F_t}$, which is much more convenient since this expectation can be estimated directly using the available sample $\widetilde{\mathbf X}_u^{(n)} \sim f_{\mathbf X_u | F_t}$, and the estimation can be more accurate.

\begin{equation} \label{eq:rewriting_exp}
    \forall u \in \mathcal P(d), \qquad \mathbb V_{f_{\mathbf X_u}} \left( \frac{f_{\mathbf X_u | F_t}(\mathbf X_u)}{f_{\mathbf X_u}(\mathbf X_u)} \right) = \mathbb E_{f_{\mathbf X_u | F_t}}\left[\frac{f_{\mathbf X_u | F_t}(\mathbf X_u)}{f_{\mathbf X_u}(\mathbf X_u)} \right] - 1.
\end{equation}

The proof of \eqref{eq:rewriting_exp} is in \ref{appendix:proofs}. Combining \eqref{eq:rewriting_var} with \eqref{eq:rewriting_exp}, the closed target Sobol indices $\text{T-S}_u^{\text{c}}$ for any $u \in \mathcal P(d)$ can then be written as

\begin{equation}\label{eq:new_writing_sobol}
    \forall u \in \mathcal P(d), \qquad \text{T-S}_u^{\text{c}} = \frac{p_t}{1-p_t} \left( \mathbb E_{f_{\mathbf X_u | F_t}} \left[\frac{f_{\mathbf X_u | F_t}(\mathbf X_u)}{f_{\mathbf X_u}(\mathbf X_u)} \right] - 1 \right).
\end{equation}

In what follows, the above expectation $\mathbb E_{f_{\mathbf X_u | F_t}}\left[\frac{f_{\mathbf X_u | F_t}(\mathbf X_u)}{f_{\mathbf X_u}(\mathbf X_u)} \right]$ is denoted by $\mathrm{E}_u$.

Following the writing obtained in \eqref{eq:new_writing_sobol}, the main quantity of interest is the density $f_{\mathbf X_u | F_t}$. Indeed, the new representation makes it possible to extend the estimation to higher dimensions, while existing estimation schemes are limited by the use of nearest-neighbor. In our case, if we are able to estimate the density $f_{\mathbf X_u | F_t}$ for large dimensional groups $u \in \mathcal P(d)$, then it becomes possible to estimate $\text{T-S}_u^{\text{c}}$. Hence the challenge lies in this estimation and the assessment of the estimation quality. The other quantities involved in \eqref{eq:new_writing_sobol} do not pose a major concern since $\widehat{p}_t$ is available as the reliability analysis has already been achieved; $f_{\mathbf X}$ is supposed to be known so we can derive $f_{\mathbf X_u}$; and conditionally to the density estimation, the expectation $\mathrm{E}_u$ can be estimated without bias with the failing $N$-sample $\widetilde{\mathbf X}^{(n)}_u \sim f_{\mathbf X_u | F_t}$.

\subsection{Discussion regarding the density estimator}

Since the reformulations \eqref{eq:new_writing_sobol} may involve possibly high-dimensional conditional densities, we discuss the required criteria to choose the appropriate estimator.

As mentioned earlier, the estimation of $\text{T-S}_u^{\text{c}}$ relies primarily on the estimation of the density $f_{\mathbf X_u | F_t}$. The reformulation proposed in \eqref{eq:new_writing_sobol} and the approach developed in this paper can be applied to any approriate density estimator. However, an estimator $\widehat{f}_{\mathbf X_u | F_t}$ for $f_{\mathbf X_u | F_t}$ must satisfy several prerequisites: it must be robust face to the dimension given the size $N$ of the available failing sample; it must be evaluable so that $\widehat{f}_{\mathbf X_u | F_t} (\widetilde{\mathbf X}_u^{(n)})$ can be obtained; and finally, it must be sufficiently flexible to handle potentially complex failure domains.

Among the possible estimators are parametric approaches, such as the von Mises–Fisher distributions and Gaussian mixture models; nonparametric approaches adapted to the dimension \cite{xu2021dens}; and approaches based on generative models, such as autoregressive models \cite{larochelle2011neuralauto, uria2016neuralauto} or normalizing flows \cite{kobyzev2020nf, papamakarios2021nf}. We refer to \cite{papamakarios2019thesis} (Part I) for a more in-depth discussion regarding density estimation in high dimensions. In addition to the estimate, we would like to have information regarding its quality and its associated error. However, it is often difficult to incorporate the error induced by the density estimation, and in some cases we lack theoretical properties for the considered estimator. This lack of information regarding the quality of the density estimation constitutes the main limitation of our methodology.

\subsection{Estimation of closed target Sobol indices}

After rewriting the closed target Sobol indices $\text{T-S}^{\text{c}}_u$ and discussing the involved quantities, we propose a new estimation scheme to estimate these indices, using only the available $N$-sample of failing points $(\widetilde{\mathbf X}^{(n)}) \sim f_{\mathbf X | F_t}$. We position our approach as \textit{given-data} since we cannot use the model $\Phi$ and rely only on an available sample. Our methodology mainly consists in replacing the quantities involved in the definition of $\text{T-S}^{\text{c}}_u$ in \eqref{eq:new_writing_sobol} by their estimate: the density $f_{\mathbf X_u | F_t}$ is estimated with $\widehat{f}_{\mathbf X_u | F_t}$ and the expectation $\mathrm{E}_u$ is estimated with Monte Carlo method. To proceed, we provide below the details of the procedure, which is summarized in Algorithm \ref{algo:estim_sobol}.

We first split the available $N$-sample $(\widetilde{\mathbf X}^{(n)})$ into two sub-samples. For $\alpha \in \: ]0,1[$, we set $N_{\mathrm{dens}} = \lfloor  \alpha N \rfloor$ and $N_{\mathrm{exp}} = N - N_{\mathrm{dens}}$. The first $N_{\mathrm{dens}}$ sample points of $(\widetilde{\mathbf X}^{(n)})$ serves to estimate the conditional density $f_{\mathbf X_u | F_t}$ with any chosen density estimator $\mathcal E$, thus providing $\widehat{f}_{\mathbf X_u | F_t}$. The remaining $N_{\mathrm{exp}}$ points of $(\widetilde{\mathbf X}^{(n)})$ are used to estimate the expectation $\mathrm{E}_u$ with the Monte Carlo method, for which we can plug-in $f_{\mathbf X_u}$ and $\widehat{f}_{\mathbf X_u | F_t}$. This splitting strategy aims at avoiding the bias that can be induced by the use of the same sample points to estimate the density and evaluate it. Second, the parameter $\alpha$ serves to allocate more points either for the estimation of $f_{\mathbf X_u | F_t}$ or for the one of $\mathrm{E}_u$, leading to the following trade-off: the more points to learn the density, the better the density is estimated, but less precise is the estimation of $\mathrm{E}_u$, and conversely. Third, this estimation procedure for $\mathrm{E}_u$ may be biased because of the error on the estimation of $f_{\mathbf X_u | F_t}$, and as mentioned earlier, there is no universal theoretical guarantee on the quality of the estimation for the density. Consequently, this limitation may induce a bias for $\text{T-S}^{\text{c}}_u$.

\vspace{5pt}
\begin{algorithm} \label{algo:estim_sobol}
 \SetKwData{Left}{left}\SetKwData{This}{this}\SetKwData{Up}{up}
\SetKwFunction{Union}{Union}\SetKwFunction{FindCompress}{FindCompress}
\SetKwInOut{Input}{input}\SetKwInOut{Output}{output}
\Input{Failing $N$-sample $(\widetilde{\mathbf X}^{(n)})$, $u \in \mathcal P(d)$, density estimator $\mathcal E$, $f_{\mathbf X}$, $\widehat{p}_t$, $\alpha \in ]0,1[$}
\Output{$\widehat{\text{T-S}}{}^{\text{c}}_u$}
\BlankLine
   \uIf{$|u| = 0$}{
    $\widehat{\text{T-S}}{}^{\text{c}}_u = 0$
  }
  \uElseIf{$|u| = d$}{
    $\widehat{\text{T-S}}{}^{\text{c}}_u = 1$
  }
  \Else{
    Set $N_{\mathrm{dens}} = \lfloor  \alpha N \rfloor$ and $N_{\mathrm{exp}} = N - N_{\mathrm{dens}}$
    
    Split $(\widetilde{\mathbf X}^{(n)})$ into two samples: $(\widetilde{\mathbf X}^{(n)}_{\mathrm{dens}})_{n=1}^{N_{\mathrm{dens}}} := (\widetilde{\mathbf X}^{(n)})_{n=1}^{N_{\mathrm{dens}}}$ and $(\widetilde{\mathbf X}^{(n)}_{\mathrm{exp}})_{n=1}^{N_{\mathrm{exp}}} := (\widetilde{\mathbf X}^{(n)})_{n=N_{\mathrm{dens}} + 1}^{N}$
    
    $\widehat{f}_{\mathbf X_u | F_t} = \mathcal E\left((\widetilde{\mathbf X}^{(n)}_{u, \mathrm{dens}})\right)$
    
    $\widehat{\mathrm{E}}_{u} = \frac{1}{N_{\mathrm{exp}}} \sum_{n=1}^{N_{\mathrm{exp}}} \frac{\widehat{f}_{\mathbf X_u | F_t}\big(\widetilde{\mathbf X}^{(n)}_{u, \mathrm{exp}}\big)}{f_{\mathbf X_u}\big(\widetilde{\mathbf X}^{(n)}_{u, \mathrm{exp}}\big)}$
    
    $\widehat{\text{T-S}}{}^{\text{c}}_u = \frac{\hat{p}_t}{1 - \hat{p}_t} (\widehat{\mathrm{E}}_{u} - 1)$
  }
 \caption{Estimation of $\text{T-S}^{\text{c}}_u$ for any $u \in \mathcal P(d)$}
\end{algorithm}

\subsection{Aggregation procedures} \label{sec:aggregation}

Now that it is possible to obtain an estimate $\widehat{\text{T-S}}{}^{\text{c}}_u$ of $\text{T-S}^{\text{c}}_u$ for $u \in \mathcal P(d)$, the target Shapley effects can be estimated through several procedures. The most direct approach is to use \eqref{eq:shap_subset}, which consists in computing $\widehat{\text{T-S}}{}^{\text{c}}_u$ for all $u \in \mathcal P(d)$. However the required number of target Sobol indices to estimate increases exponentially with the dimension as mentioned earlier and estimating all of them becomes rapidly intractable. The other mentioned approach consists in rewriting the Shapley effects as an expectation over the set of permutations $\mathfrak S_d$, following \eqref{eq:shap_perm}. This expectation can be estimated by crude Monte Carlo with a $M$-sample of permutations $(\pi_m)_{m=1}^M$ uniformly drawn from $\mathfrak S_d$, hence providing the following estimate for the target Shapley effect $\text{T-Sh}_i$ associated to $X_i$

\begin{equation}\label{eq:estim_shap_perm}
    \widehat{\text{T-Sh}}_i = \frac{1}{M} \sum_{m=1}^M \widehat{\text{T-S}}{}^{\text{c}}_{P_i(\pi_m) \, \cup \, \{i\}} - \widehat{\text{T-S}}{}^{\text{c}}_{P_i(\pi_m)},
\end{equation}

where $P_i(\pi)$ for $\pi \in \mathfrak S_d$ has been defined in \eqref{eq:shap_perm}. This procedure provides a trade-off between reachable precision and computational cost (in terms of post-processing) \cite{maleki2013boundinderror} as it induces an additional variability compared with the subset aggregation \eqref{eq:shap_subset}. Furthermore, it should be noted that the crude Monte Carlo estimations for both the aggregation procedure and the estimation of $\text{E}_u$ do not introduce any additional bias into our procedure. However, any potential bias introduced by the estimation of the density $f_{\mathbf X_u | F_t}$ is reflected in the final estimate of the target Shapley effect.

The naive approach of \eqref{eq:estim_shap_perm} with the groups of variables $P_i(\pi) \, \cup \, \{i\}$ and $P_i(\pi)$ for a permutation $\pi$ leads to some duplicate calculations. To circumvent this computational flaw, we rely on the approach named \texttt{ApproShapley} \cite{song2016shapley} in order to avoid these duplicates, which allows to reduce the number of required Sobol indices to estimate to $M(d-1)$. We provide below the main methodology and refer to the associated paper \cite{song2016shapley} for the details.

Consider the following example, where $d=4$ and let $\pi$ be the permutation $\pi = (2,1,4,3)$. We have $P_{\pi(2)}(\pi) = P_{1}(\pi) = \{2\}$ and $P_{\pi(3)}(\pi) = P_{4}(\pi) = \{2,1\} = P_{\pi(2)}(\pi) \cup \{\pi(2)\}$. Hence the marginal contribution of the variable $X_{\pi(2)} = X_1$ for the order $\pi$ is

\begin{equation} \label{eq:song_comput}
    \text{T-S}^{\text{c}}_{\{2,1\}} - \text{T-S}^{\text{c}}_{\{2\}} = \text{T-S}^{\text{c}}_{P_{\pi(2)}(\pi) \cup \{\pi(2)\}} - \text{T-S}^{\text{c}}_{P_{\pi(2)}(\pi)} =  \text{T-S}^{\text{c}}_{P_{\pi(3)}(\pi)} - \text{T-S}^{\text{c}}_{P_{\pi(2)}(\pi)}.
\end{equation}

Reasoning in the same way for the variable $X_{\pi(3)} = X_4$, we obtain

\begin{equation} \label{eq:song_comput2}
    \text{T-S}^{\text{c}}_{\{2,1,4\}} - \text{T-S}^{\text{c}}_{\{2,1\}} = \text{T-S}^{\text{c}}_{P_{\pi(4)}(\pi)} - \text{T-S}^{\text{c}}_{P_{\pi(3)}(\pi)}.
\end{equation}

We remark that the quantity $\text{T-S}^{\text{c}}_{P_{\pi(3)}(\pi)}$ is used in both \eqref{eq:song_comput} and \eqref{eq:song_comput2}. The approach \texttt{ApproShapley} keeps the estimation of $\text{T-S}^{\text{c}}_{P_{\pi(3)}(\pi)}$ already made for \eqref{eq:song_comput} and will reuse it for \eqref{eq:song_comput2}. Hence, we only need to compute the quantities $\text{T-S}^{\text{c}}_{P_{\pi(i)}(\pi)}$ for $1 < i \leq d$ in order to compute the marginal contributions of the $d$ inputs variables for an order $\pi$, which corresponds to $d-1$ Sobol indices for each permutation.

To summarize our estimation procedure, we have proposed a new rewriting of the closed target Sobol indices $\text{T-S}_u^c$ for any $u \in \mathcal P(d)$, which allows to circumvent the problem of the dimension through the introduction of a conditional density of potentially large dimensions and which can be estimated with any appropriate estimator. Once this density has been estimated, we then estimate the expectation $\mathrm{E}_u$ using Monte Carlo sampling, and then we can obtain an estimate of the closed target Sobol $\text{T-S}_u^c$. The target Shapley effects can then be estimated by drawing a sample of $M$ permutations, and by applying our aforementioned method to estimate each of the required target Sobol indices.

\section{Error quantification} \label{sec:err_quantif}

The methodology described in the previous Section \ref{sec:targ_shap_nf} implies a three-step estimation: a density estimation for $f_{\mathbf X_u | F_t}$, a MC estimation for $\text{E}_u$ and another MC estimation for $\text{T-Sh}_i$. These three steps each introduce an error in the final estimation, and we propose in this section to empirically quantify these three errors only with the available failing $N$-sample $(\widetilde{\mathbf X}^{(n)})_{n=1}^N$. This constitutes a new contribution and a major benefit for practitioners as it does not require the estimation to be repeated and therefore it does not require any additional call to the model $\Phi$.

\subsection{Error estimation for the Sobol indices} \label{sec:err_sobol}

To proceed, we first estimate the error for the closed target Sobol index $\text{T-S}_u^c$, for any $u \in \mathcal P(d)$. As mentioned previously, it is difficult to isolate the respective error caused by the density estimation, and even if it would not be the case, we would be constrained to the chosen estimator. Instead, we propose an empirical approach that is valid for any density estimator. Moreover, we propose to jointly quantify the errors of $f_{\mathbf X_u | F_t}$ and $\text{E}_u$ for $\text{T-S}_u^c$ instead of trying to isolate each. This procedure, summarized in Algorithm \ref{algo:uq_sobol} below, produces $J \in \mathbb N^*$ estimates of $\text{T-S}_u^c$ though a resampling approach. The idea is to change at each repetition the sub-sample used to estimate the density and the sub-sample used to estimate the expectation.

To be more precise, the indices $(1, \ldots, N)$ are permuted with permutation $\delta \in \mathfrak S_N$ into $(\delta(1), \ldots, \delta(N))$ to obtain the $N$-sample $(\widetilde{\mathbf X}^{(\delta(n))})_{n=1}^N$. Then, applying Algorithm \ref{algo:estim_sobol} with the permuted sample $(\widetilde{\mathbf X}^{(\delta(n))})$ provides an estimation of $\text{T-S}_u^c$ different from the one with the initial order since different sample points are used to estimate the density and the expectation. Repeating this procedure $J$ times allows us to obtain an estimate of the error introduced by both the density estimation and the MC estimation in the estimate of the closed target Sobol indices.

\vspace{5pt}
\begin{algorithm}[H] \label{algo:uq_sobol}
 \SetKwData{Left}{left}\SetKwData{This}{this}\SetKwData{Up}{up}
\SetKwFunction{Union}{Union}\SetKwFunction{FindCompress}{FindCompress}
\SetKwInOut{Input}{input}\SetKwInOut{Output}{output}
\Input{Failing $N$-sample $(\widetilde{\mathbf X}^{(n)})$, $u \in \mathcal P(d)$, density estimator $\mathcal E$, $f_{\mathbf X}$, $\widehat{p}_t$, proportion $\alpha \in ]0,1[$, $J \in \mathbb N^*$}
\Output{a set of $J$ estimations for the Sobol index $\text{T-S}_u^c$}
\BlankLine
    \For{$j=1,2,\ldots, J$}{
    Draw $\delta_j$ uniformly from the set of permutations $\mathfrak S_N$

    $\widehat{\text{T-S}}{}^{\text{c}, (j)}_u = \textsf{Algorithm1}((\widetilde{\mathbf X}^{\delta_j(n)}), u, \mathcal E, f_{\mathbf X}, \widehat{p}_t, \alpha)$
    }
 \caption{Uncertainty quantification for $\text{T-S}^{\text{c}}_u$}
\end{algorithm}
\vspace{5pt}

\subsection{Error estimation for the marginal contributions}

Once the uncertainty quantification on the closed target Sobol indices have been made possible, we can reuse this procedure to quantify the uncertainty on the marginal contributions detailed in Section \ref{sec:aggregation}. Since the target Shapley effects are averages of marginal contributions as in \eqref{eq:shap_perm}, having access to the marginal contributions' uncertainty will provide the uncertainty for the target Shapley effects as detailed in Section \ref{sec:err_permutations}.

Recall that for a permutation $\pi \in \frak S_d$, the marginal contribution of the input variable $X_{i}$ is given by

\begin{equation}\label{eq:def_marg_contrib}
    mc(i, \pi) = \text{T-S}^{\text{c}}_{P_{i}(\pi) \cup \{i\}} - \text{T-S}^{\text{c}}_{P_{i}(\pi)}.
\end{equation}

The estimation of $mc(i, \pi)$ requires the estimation of both $\text{T-S}^{\text{c}}_{P_{i}(\pi) \cup \{i\}}$ and $\text{T-S}^{\text{c}}_{P_{i}(\pi)}$. Consequently, to obtain the uncertainty on $mc(i, \pi)$, we propose to apply the previous Algorithm \ref{algo:uq_sobol} for these two Sobol indices, which will provide $J$ estimates for each. Then, we independently sample one of their $J$ estimates for each Sobol index and we take the difference, which provide an estimate of the marginal contribution $mc(i, \pi)$. Repeating this last step $L \in \mathbb N^*$ times provides $L$ estimates of the marginal contribution $mc(i, \pi)$, which we may write as $\widehat{mc}(i, \pi, \ell)$, $1 \leq \ell \leq L$ to emphasize the repetition. This final step provides the resulting uncertainty on $mc(i, \pi)$.

Now, following again Section \ref{sec:aggregation}, we instead consider the \texttt{ApproShapley} variant in order to reduce the computational cost. This means that we consider the marginal contribution of the input variable $X_{\pi(i)}$ for the order $\pi$, given by

\begin{align} \label{eq:def_marg_contrib_song}
    mc(\pi(i), \pi) &= \text{T-S}^{\text{c}}_{P_{\pi(i)}(\pi) \cup \{\pi(i)\}} - \text{T-S}^{\text{c}}_{P_{\pi(i)}(\pi)} \\ 
    &= \text{T-S}^{\text{c}}_{P_{\pi(i+1)}(\pi)} - \text{T-S}^{\text{c}}_{P_{\pi(i)}(\pi)},
\end{align}
where the last equality holds for $ 1 \leq i < d$ as seen in \eqref{eq:song_comput}, and $\text{T-S}^{\text{c}}_{P_{\pi(d)}(\pi) \cup \{\pi(d)\}} = \text{T-S}^{\text{c}}_{[d]} = 1$.

We then apply the aforementioned procedure to recover $L$ estimates of $mc(\pi(i), \pi)$, by applying Algorithm \ref{algo:uq_sobol} to both $\text{T-S}^{\text{c}}_{P_{\pi(i+1)}(\pi)}$ and $\text{T-S}^{\text{c}}_{P_{\pi(i)}(\pi)}$. Hence, for a permutation $\pi$ and an input variable $X_{\pi(i)}$, we recover $L$ estimates of the corresponding marginal contribution. Since \texttt{ApproShapley} estimates sequentially $mc(\pi(i), \pi)$ for $1 < i \leq d$, we obtain $L$ estimates for each $X_{\pi(i)}$. Finally, since we consider a $M$-sample of permutations $(\pi_m)$ uniformly drawn from $\frak S_d$ to estimate the target Shapley effects $\text{T-Sh}_i$, we need to obtain the uncertainty for $mc(\pi_m(i), \pi_m)$ for all $1 \leq m \leq M$. Consequently, we apply the procedure for each permutation $\pi_m$, providing $\widehat{mc}(i, \pi_m, \ell)$, $1 \leq \ell \leq L$. Algorithm \ref{algo:uq_marg_contri} below summarizes the procedure and provides the technical details.

\vspace{5pt}
\begin{algorithm}[H] \label{algo:uq_marg_contri}
 \SetKwData{Left}{left}\SetKwData{This}{this}\SetKwData{Up}{up}
\SetKwFunction{Union}{Union}\SetKwFunction{FindCompress}{FindCompress}
\SetKwInOut{Input}{input}\SetKwInOut{Output}{output}
\Input{$M \in \mathbb N^*$, $L \in \mathbb N^*$ and $J \in \mathbb N^*$, $(\widetilde{\mathbf X}^{(n)})$, $\mathcal E$, $f_{\mathbf X}$, $\widehat{p_t}$, $\alpha$}
\Output{a $3$-dimensional array $\widehat{mc}$ of size $d \times M \times L$}
\BlankLine
    Get $M$-sample $(\pi_m)$ where $\pi_m \sim \mathcal U(\mathfrak S_d)$ \;
    \For{$m=1,2,\ldots, M$}{
        Set $\mathrm{Prev} = (0)_{j=1}^J$ \hspace{1cm} \textsf{$\backslash\backslash$ a vector of $0$'s, of size $J$}

        \For{$i=1,2,\ldots, d$}{
            \uIf{$i = d$}{
                Set $\widehat{\mathbf{ts}}_i = (1)_{j=1}^J$ \hspace{1cm} \textsf{$\backslash\backslash$ a vector of $1$'s, of size $J$}
            }
            \uElse{
                Set $\widehat{\mathbf{ts}}_i = \textsf{Algorithm2}((\widetilde{\mathbf X}^{(n)}), P_{\pi_m(i+1)} (\pi_m), \mathcal E, f_{\mathbf X}, \widehat{p}_t, \alpha, J)$ \hspace{0.1cm} \textsf{$\backslash\backslash$ $J$ estimates of} $\text{T-S}^c_{P_{\pi_m(i+1)}(\pi_m)}$
            }
            \For{$\ell=1,2,\ldots, L$}{
                Sample $\mathfrak{J}_{1,\ell}$ and $\mathfrak{J}_{2,\ell}$ from $\mathcal U\{1, \ldots, J\}$

                Set $\widehat{mc}(\pi_m(i), m, \ell) := \widehat{\mathbf{ts}}_i^{(\mathfrak{J}_{1,\ell})} - \mathrm{Prev}^{(\mathfrak{J}_{2,\ell})}$
            }
            $\mathrm{Prev} = \widehat{\mathbf{ts}}_i$ 
        }
    }
 \caption{Uncertainty quantification for marginal contributions}
\end{algorithm}
\vspace{5pt}

\subsection{Error estimation for the Shapley effects}\label{sec:err_permutations}

To conclude our quantification of the error, we propose a probabilistic error bound that provides the probability of making an error less than $\varepsilon_i$ for the target Shapley effect $\widehat{\text{T-Sh}}_i$. More specifically, we reuse the uncertainty obtained for the marginal contributions in Algorithm \ref{algo:uq_marg_contri} to ensure that $|\widehat{\text{T-Sh}}_i - \mathbb E\big[\widehat{\text{T-Sh}}_i\big]| \leq \varepsilon_i$ with a probability at least $1 - \delta$ for $\varepsilon_i, \delta > 0$, or equivalently

\begin{equation} \label{eq:prob_bound}
    \mathbb P\left(\big|\widehat{\text{T-Sh}}_i - \mathbb E\big[\widehat{\text{T-Sh}}_i\big]\big| \geq \varepsilon_i\right) \leq \delta,
\end{equation}

where $\mathbb E\big[\widehat{\text{T-Sh}}_i\big] = \mathbb E_{\pi_m, \ell}\big[\widehat{\text{T-Sh}}_i\big]$ with $\pi_m \sim \mathcal U(\mathcal S(d))$ and $\ell \sim \mathcal U(\{1, \ldots, L\})$. To obtain this result, we rely on the work of \cite{maleki2013boundinderror}, which uses Hoeffding’s inequality, which we recall here. Let $V_1, \ldots, V_M$ be independent real random variables such that each $V_m$ takes values in $[a_m, b_m]$, where $a_m$ and $b_m$ are real numbers, and let $r_m = b_m - a_m$ denote the range of $V_m$. Let $S = V_1 + \ldots + V_M$. Then, for any $\varepsilon > 0$, we have

\begin{equation} \label{eq:hoeffding_ineq}
    \mathbb P\left(\left|S - \mathbb E\left[S\right]\right| \geq \varepsilon\right) \leq 2\exp\left(- \frac{2\varepsilon^2}{\sum_{m=1}^M r_m^2} \right).
\end{equation}

or equivalently

\begin{equation} \label{eq:hoeffding_ineq_equiv}
    \mathbb P\left(\left|\frac{S}{M} - \mathbb E\left[\frac{S}{M}\right]\right| \geq \varepsilon\right) \leq 2\exp\left(- \frac{2\varepsilon^2 M^2}{\sum_{m=1}^M r_m^2} \right).
\end{equation}

This last inequality bounds the probability that a random variable $\frac{1}{M} \sum_{m=1}^M V_m$ deviates from its expected value for a fixed number $M$. We can now draw an analogy with the target Shapley effects $\widehat{\text{T-Sh}}_i$ by setting $V_m = \widehat{mc}(i, \pi_m)$, since $\widehat{\text{T-Sh}}_i = \frac{1}{M} \sum_{m=1}^M \widehat{mc}(i, \pi_m)$. Here we make the assumption that the marginal contributions $\widehat{mc}(i, \pi_m), 1 \leq m \leq M$ are independent to apply Hoeffding's inequality. Moreover, we need to determine the range $r_m$ of $\widehat{mc}(i, \pi_m)$. To do this, we can estimate it using the output of Algorithm \ref{algo:uq_marg_contri}, which consists of $L$ estimates $\widehat{mc}(i, \pi_m, \ell)$, $1 \leq \ell \leq L$. An estimate $\widehat{r}_m$ is obtained by considering

\begin{equation} \label{eq:range_pi_m}
    \widehat{r}_m = \max_{1 \leq \ell \leq L} \widehat{mc}(i, \pi_m, \ell) - \min_{1 \leq \ell \leq L} \widehat{mc}(i, \pi_m, \ell).
\end{equation}

Hence, we obtain an estimate of the range of marginal contributions specific to the permutation $\pi_m$. In addition, with the $L$ estimates $\widehat{mc}(i, \pi_m, \ell)$ from Algorithm \ref{algo:uq_marg_contri}, we compute $\widehat{mc}(i, \pi_m) = \frac{1}{L} \sum_{\ell = 1}^L \widehat{mc}(i, \pi_m, \ell)$. Furthermore, we would like to simplify by considering a universal range for $X_i$, denoted $r_{(i)}$, which would allow us to bound all marginal contributions for the variable $X_i$, regardless of the permutation $\pi_m$ under consideration (this will simplify the expression $\sum_{m=1}^M r_m^2$ in Hoeffding's inequality \eqref{eq:hoeffding_ineq}). This universal range $r_{(i)}$ is estimated by setting

\begin{equation} \label{eq:range_universal_xi}
    \widehat{r}_{(i)} = \max_{1 \leq m \leq M} \max_{1 \leq \ell \leq L} \widehat{mc}(i, \pi_m, \ell) - \min_{1 \leq m \leq M} \min_{1 \leq \ell \leq L} \widehat{mc}(i, \pi_m, \ell).
\end{equation}

We remark that the range $\widehat{r}_m$ only takes into account the randomness from the $L$ repetitions (meaning the error from the estimation of $f_{\mathbf X_u | F_t}$ and $\text{E}_u$), while the range $\widehat{r}_{(i)}$ also takes into account the randomness coming from permutation sampling. Consequently, since $\widehat{r}_{(i)}$ is a suitable range for every $\widehat{mc}(i, \pi_m), 1 \leq m \leq M$, we obtain

\begin{equation} \label{eq:bound_hoeffding}
    \mathbb P\left(\big|\widehat{\text{T-Sh}}_i - \mathbb E[\widehat{\text{T-Sh}}_i]\big| \geq \varepsilon_i \right) \leq 2\exp\left(- \frac{2\varepsilon_i^2M}{\widehat{r}_{(i)}^2} \right).
\end{equation}

Finally, $\mathbb P(|\widehat{\text{T-Sh}}_i - \mathbb E[\widehat{\text{T-Sh}}_i]| \geq \varepsilon_i) \leq \delta$ is verified when

\begin{equation} \label{eq:bound_error}
    2\exp\left(- \frac{2 \, \varepsilon_i^2M}{\widehat{r}_{(i)}^2} \right) \leq \delta \qquad \Longleftrightarrow \qquad \varepsilon_i \geq \sqrt{\frac{\ln(2/\delta) \, \widehat{r}_{(i)}^2}{2 M}}.
\end{equation}

One should remark that this result does not provide information regarding the true Shapley effects $\text{T-Sh}_i$, but rather it says that our average estimation $\mathbb E[\widehat{\text{T-Sh}}_i]$ lies in the interval $[\widehat{\text{T-Sh}}_i - \varepsilon_i, \widehat{\text{T-Sh}}_i + \varepsilon_i]$ with probability greater than $1 - \delta$. Indeed we do not necessarily have $\mathbb E[\widehat{\text{T-Sh}}_i] = \text{T-Sh}_i$ due to the bias from the density estimation $\widehat{f}_{\mathbf X_u | F_t}$.

\subsection{Practical guidances}

In this section, we provide some pratical guidances for the estimation scheme as well as the expected computational cost based on the parameters under consideration. First of all, for a fixed number of permutations $M$, a number of $M \times (d-1) \times J$ estimations of target Sobol indices is required according to Algorithm \ref{algo:uq_marg_contri}. Thus, if estimating a target Sobol index takes an average of $\tau$ seconds, then the expected computation time is of order $M \times (d-1) \times J \times \tau$ seconds. To determine the time $\tau$ for a given density estimator, we propose the following approach: for several groups $u \in \mathcal P(d)$ and with the available failing sample size, one can test different values of $\alpha \in [0,1]$ to determine the most appropriate value. Once the parameter $\alpha$ has been set, the practitioner can test several groups $u \in \mathcal P(d)$ of different sizes and take $\tau$ as the average of the associated times. This will yield an average computation time for the target Sobol indices using the associated density estimator. Finally, setting $J=10$ and $L=10$ should provide sufficient information on the uncertainty, but the practitioner is free to adjust this parameter to verify the value in question. Note that we omit the parameter $L$ in the final computational cost as its impact is negligible in Algorithm \ref{algo:uq_marg_contri}.

The values of $N$, $\alpha$, $J$ and $L$ being fixed, we can determine the required number of permutations to reach a given precision level. Remark from Section \ref{sec:err_permutations} that if the universal range $r_{(i)}$ associated to $X_i$ is small, then only few permutations are required to reach a small error for this specific input variable and on the contrary if $r_{(i)}$ is large, then a large number of permutations will be required to reach a small error. Suppose that a maximal error $\varepsilon_\star$ is required for all the target Shapley effects, and note $r_{\text{max}} := \max_{1 \leq i \leq d} r_{(i)}$ the maximal range. Then from \eqref{eq:bound_error} the minimum number of permutations $M_\star$ required to reach the precision $\varepsilon_\star$ for all the target Shapley effects is $M_\star \geq \left\lceil\frac{\ln(2/\delta) r_{\text{max}}^2}{2 \varepsilon_\star^2} \right\rceil$. As a rule of thumb, one can apply the aforementioned approach with $M=100$ permutations to have an estimation of the ranges $r_{(i)}$ for $1 \leq i \leq d$, then compute $r_{\text{max}}$ and recover $M_\star$. Then the procedure is applied again with the remaining number of required permutation $\max(M_\star - M, 0)$ and the $M_\star$ marginal contributions are all gathered. Therefore, the expected computation cost is of order $M_\star \times (d-1) \times J \times \tau$ seconds for the estimation of $d$ target Shapley effects.

\section{Numerical applications} \label{sec:num_appli}

In this section, we propose to illustrate our methodology on three use cases. First, we consider an analytical Gaussian linear model for which we illustrate how the different quantities involved in our procedure impact the estimation of the closed target Sobol indices and of the target Shapley effects. Second, we consider the fire-spread model, a real world situation with a more complex input distribution and black-box. Finally, we propose to apply our methodology to monitor the impact of the input variables during each step of an adaptive importance sampling scheme.

All the computer codes and data can be found following the link

\begin{center}\label{repo}
    \href{https://github.com/lucasmonteiro4/target-shapley-density}{\textsf{https://github.com/lucasmonteiro4/target-shapley-density}},
\end{center}

and the machine used for the simulations and the computations is a desktop Windows 11 64 bit, processor i5 13th Gen, 16Gb RAM.

\subsection{Gaussian Linear case} \label{sec:app_gl}

The Gaussian linear case is a simple analytical model for which the function $\Phi$ is defined by $$\Phi(\mathbf X) = \bm\beta^\top \mathbf X = \beta_1X_1 + \beta_2 X_2 + \ldots + \beta_d X_d \, ,$$ where $\bm\beta \in \mathbb R^d \backslash \{\mathbf 0_d\}$ is a vector of coefficients representing the weight of each input variable in the linear combination. Moreover, $\mathbf X \sim \mathcal N(\bm\mu, \bm\Sigma)$ where $\bm\mu \in \mathbb R^d$ and $\bm\Sigma$ is a $d \times d$ symmetric positive-definite matrix with real-valued coefficients. In our context, $\bm\Sigma$ is not necessarily diagonal or block-diagonal. For a given failure threshold $t \in \mathbb R$, the theoretical values for the closed target Sobol indices can be computed, and consequently we have access to reference values for the theoretical target Shapley effects. We refer to the existing literature for the expression of the theoretical values \cite{ilidrissi2021shap, demangechryst2023shaprosa}. For our example, we set the dimension to $d=30$, $\beta_2 = \beta_8 = \beta_{11} = 0$, $\beta_9 = \beta_{15} = 5$, $\beta_{17} = 10$, $\beta_4 = \beta_{29} = 15$, $\beta_{24} = 20$ and all other values of $\bm\beta$ are set to $1$. Therefore, we should observe that $X_{4}, X_{24}$ and $X_{29}$ have a large impact on the variance. We set $\bm\mu = \bm 0$ and since $\bm\Sigma$ is too big to be included in the Appendix, we specified it in the \href{https://github.com/lucasmonteiro4/target-shapley-density/tree/main/gaussian_linear}{Github repository}. The threshold $t$ is set to $105$, so as to obtain $p_t \approx 10^{-3}$.

\begin{center}
    \begin{figure}[h!]
    \centering
    \hspace{-20pt}
    \begin{minipage}{.3\textwidth}
        \centering
        \includegraphics[width=5.5cm]{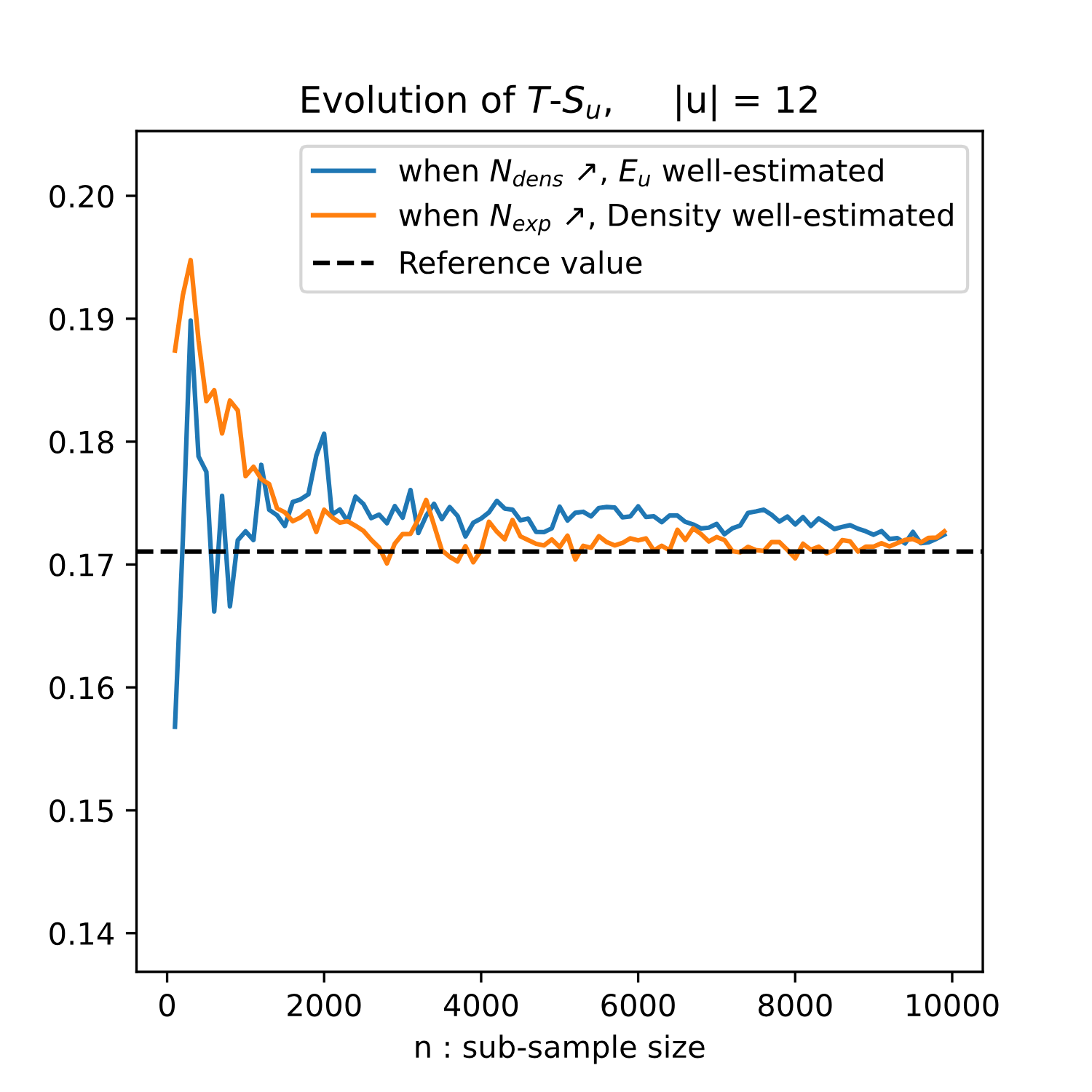}
        {\footnotesize $(a)$ \\ $|u| = 12$}
    \end{minipage}
    \hspace{10pt}
    \begin{minipage}{0.3\textwidth}
        \centering
        \includegraphics[width=5.5cm]{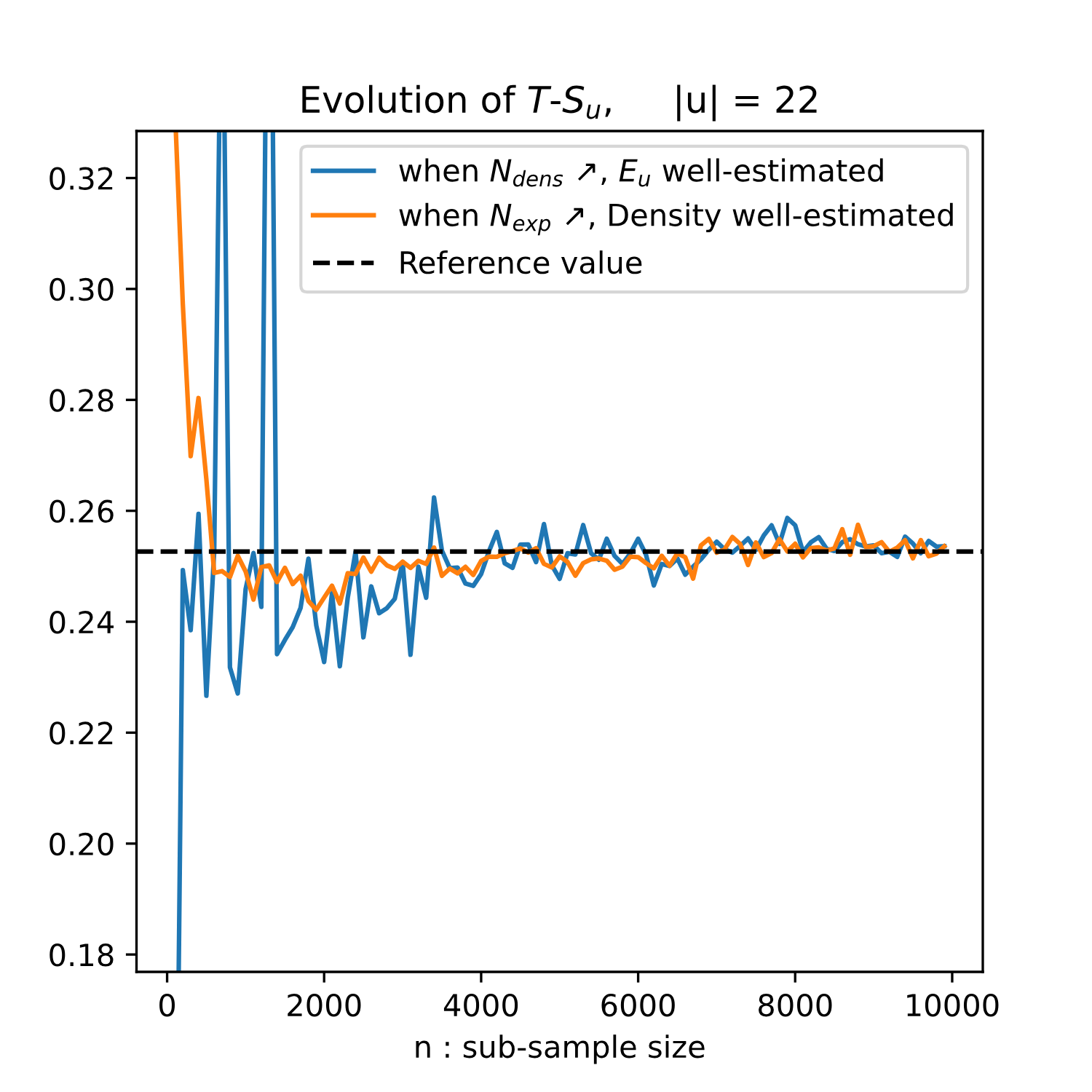}
        {\footnotesize $(b)$ \\ $|u| = 22$}
    \end{minipage}
    \hspace{10pt}
    \begin{minipage}{0.3\textwidth}
        \centering
       \includegraphics[width=5.5cm]{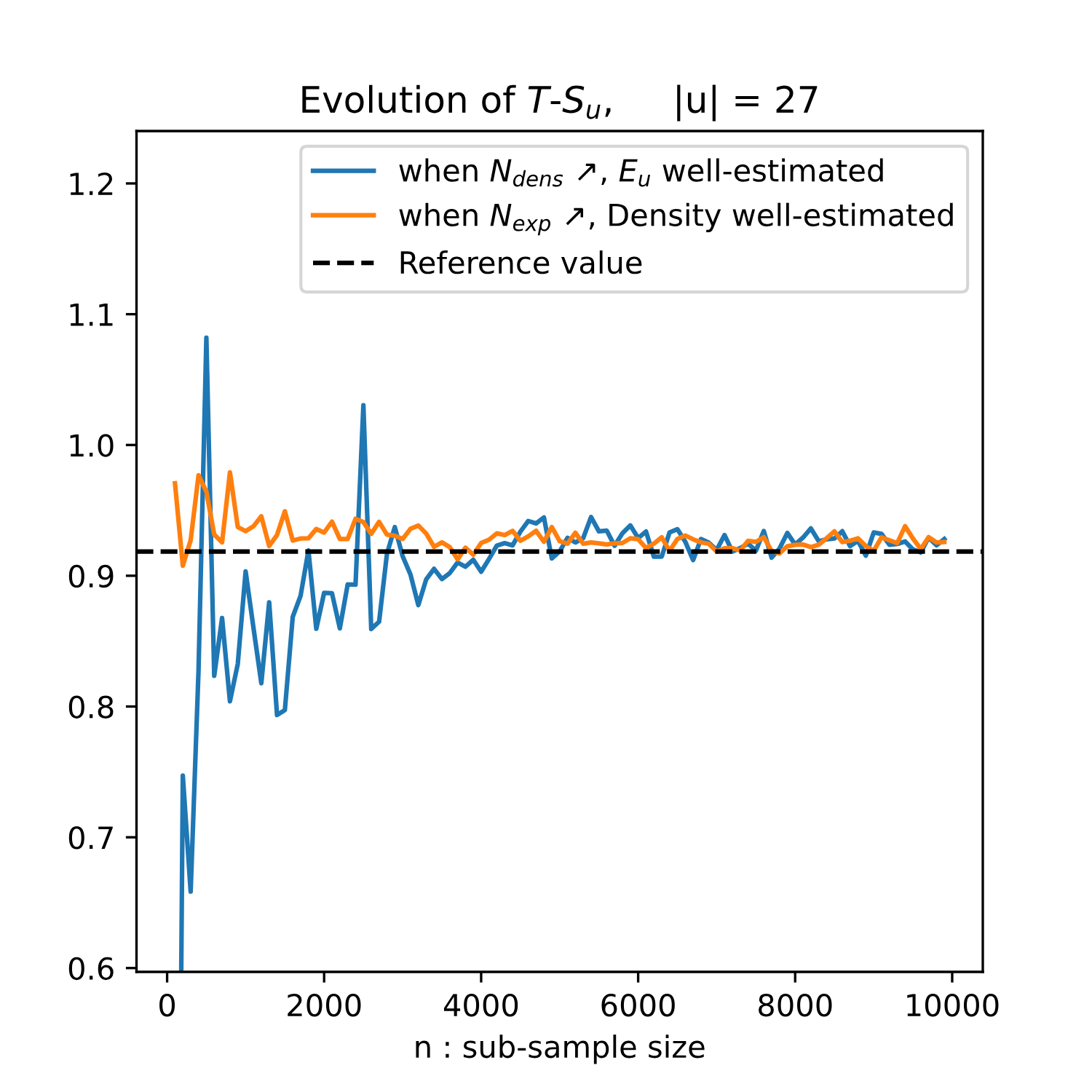}
        {\footnotesize $(c)$ \\ $|u| = 27$}
    \end{minipage}
    \caption{Evolution of the estimate of the closed target Sobol indices when the expectation $\text{E}_u$ is well estimated and the estimate of the density $f_{\mathbf X_u | F_t}$ is gradually improved (in blue) \textit{vs.} when the density is well estimated and the estimate of the expectation is gradually improved (in orange).}
    \label{fig:fig1_2estim}
\end{figure}
\end{center}

\begin{center}
    \begin{figure}[h!]
    \centering
    \hspace{-20pt}
    \begin{minipage}{.3\textwidth}
        \centering
        \includegraphics[width=5.5cm]{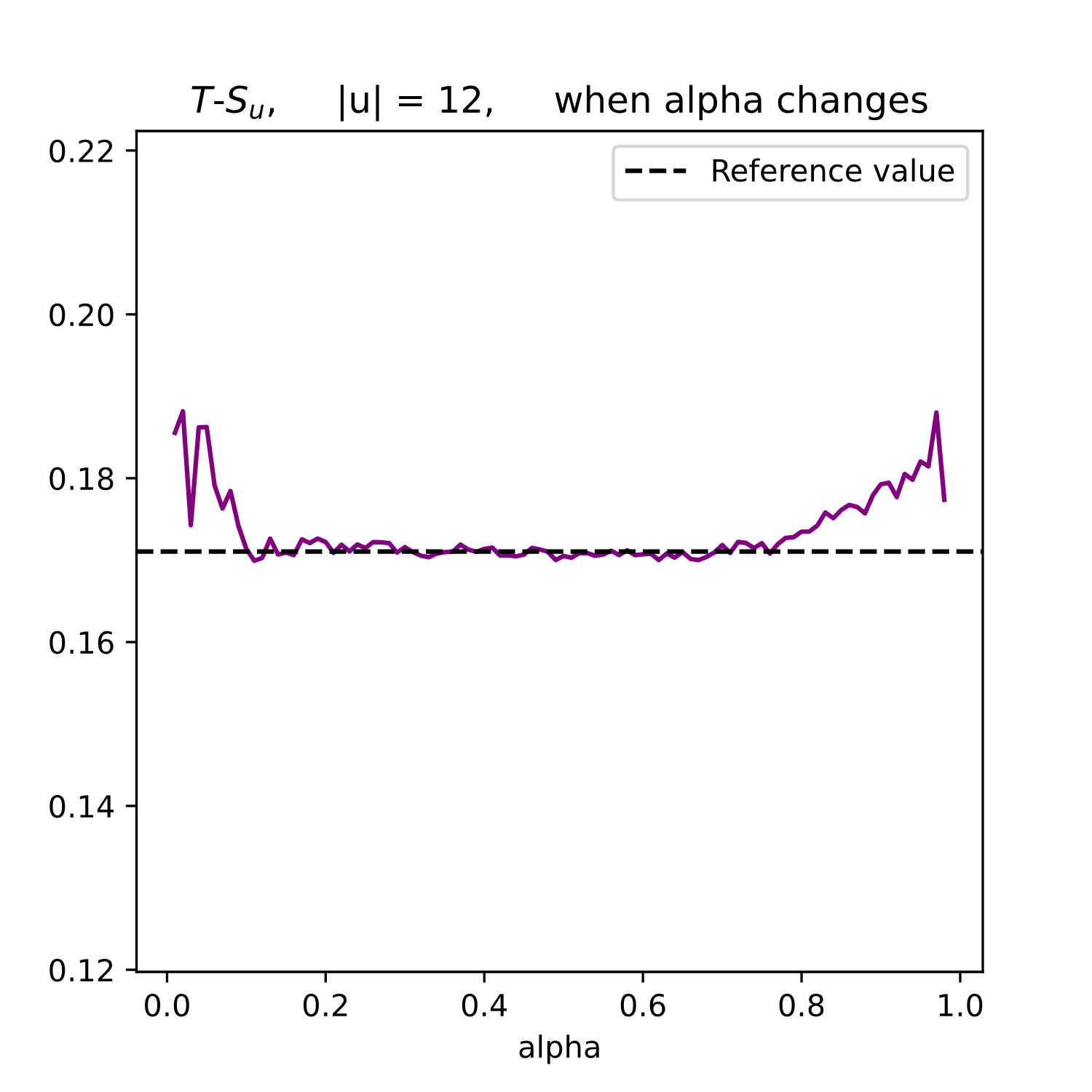}
        {\footnotesize $(a)$ \\ $|u| = 12$}
    \end{minipage}
    \hspace{10pt}
    \begin{minipage}{0.3\textwidth}
        \centering
        \includegraphics[width=5.5cm]{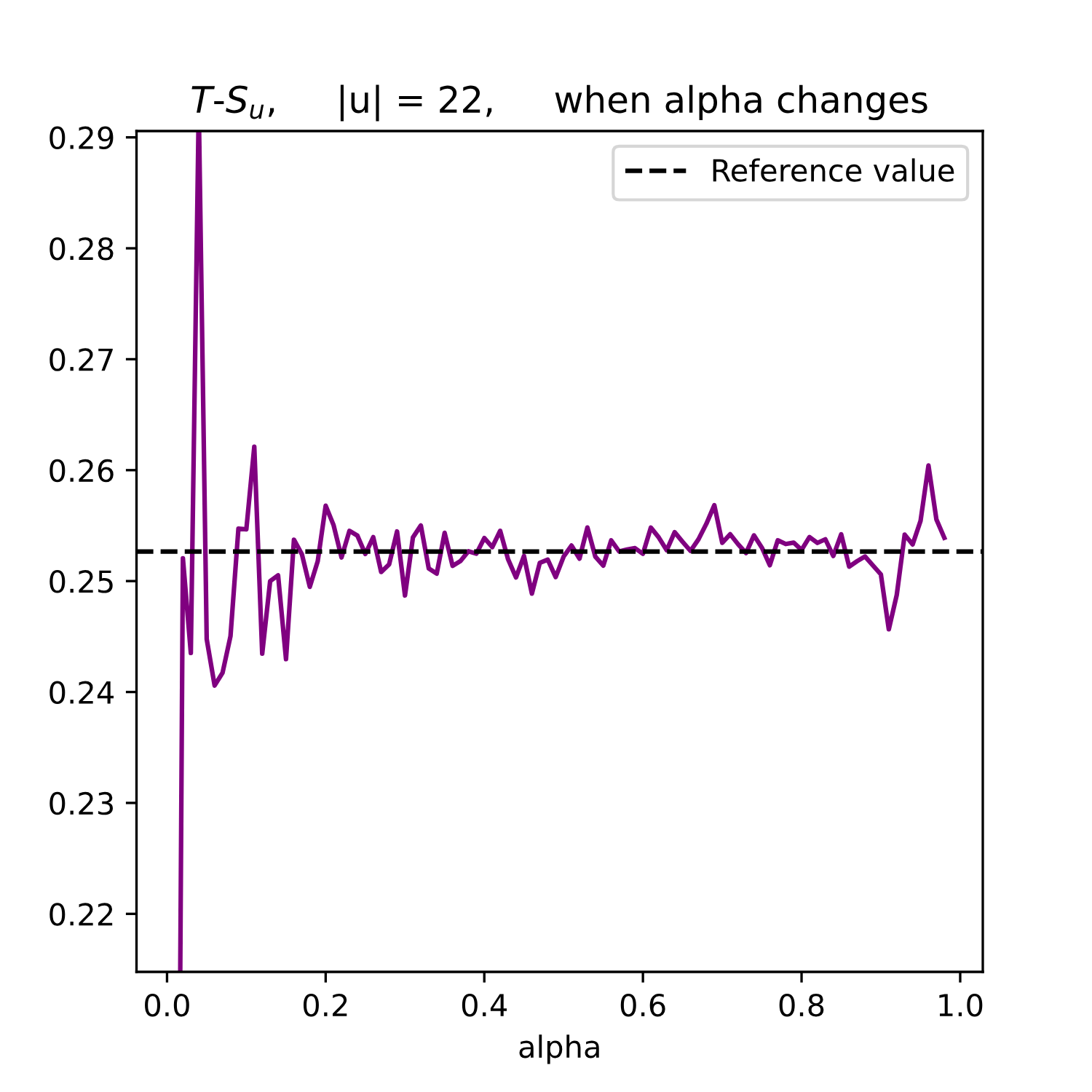}
        {\footnotesize $(b)$ \\ $|u| = 22$}
    \end{minipage}
    \hspace{10pt}
    \begin{minipage}{0.3\textwidth}
        \centering
       \includegraphics[width=5.5cm]{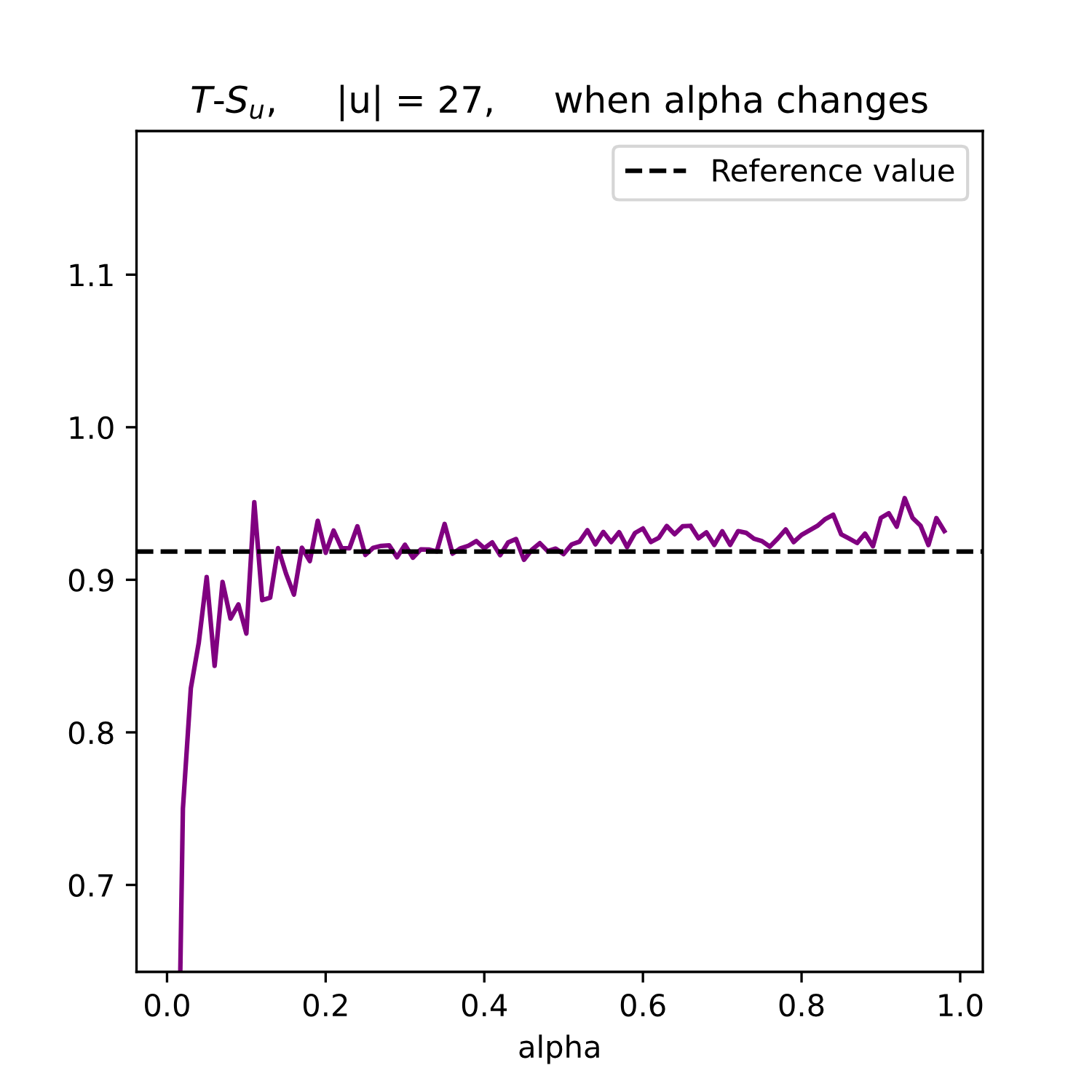}
        {\footnotesize $(c)$ \\ $|u| = 27$}
    \end{minipage}
    \caption{Estimation of the closed target Sobol indices when the parameter $\alpha \in [0,1]$ changes.}
    \label{fig:fig2_alpha}
\end{figure}
\end{center}

First of all, we study how the various quantities involved in the estimation of the closed target Sobol $\text{T-S}_u$ affect the quality of that estimate. To proceed, we examine the role played by the failing sample size $N$, the density $f_{\mathbf X_u | F_t}$, the expectation $\text{E}_u$, and the parameter $\alpha$. First, we illustrate in Figure \ref{fig:fig1_2estim} how the estimation of $\text{T-S}_u$ is sensitive to the estimation of the quantities $f_{\mathbf X_u | F_t}$ and $\text{E}_u$. We use a failing sample of size $N = 20 \ 000$ and carry out two analyses. First, we (randomly) allocate $N_{\text{exp}} = 10 \ 000$ points to estimate $\text{E}_u$ – which yields a good estimate – and to estimate the density $f_{\mathbf X_u | F_t}$ we allocate $N_{\text{dens}} \in \Iintv{1, 10 \ 000}$ points from the remaining sample, then estimate the associated Sobol $\text{T-S}_u$. Few points for the estimation of $f_{\mathbf X_u | F_t}$ should provide a poor estimate and increasing the number of points should improve that estimate. This is represented by the blue curve in Figure \ref{fig:fig1_2estim}. We also consider the reverse and allocate $N_{\text{dens}} = 10 \ 000$ points to the estimate of the density $f_{\mathbf X_u | F_t}$ - which yields a correct estimate - and to estimate the expectation $\text{E}_u$ we allocate $N_{\text{exp}} \in \Iintv{1, 10 \ 000}$ points from the remaining sample, then we estimate the associated Sobol index $\text{T-S}_u$. This is represented by the orange curve in Figure \ref{fig:fig1_2estim}. We also illustrate in Figure \ref{fig:fig2_alpha} the role of the parameter $\alpha$ in Algorithm \ref{algo:estim_sobol}, which represents the different partitions of the available failing sample $(\widetilde{\mathbf X}^{(n)})$, $\alpha \approx 0$ means that almost all points are used to estimate the expectation $\text{E}_u$, and few points for $f_{\mathbf X_u | F_t}$, and $\alpha \approx 1$ means the contrary. For this second analysis, we also use the failing sample of size $N = 20 \ 000$. For both Figures \ref{fig:fig1_2estim} and \ref{fig:fig2_alpha}, we illustrate three Sobol indices, one of order $12$, one of order $22$ and one of order $27$. Moreover, we use a GMM with $5$ components for the estimation of the density $f_{\mathbf X_u | F_t}$ for the three considered $u \in \mathcal P(d)$.

\begin{center}
    \begin{figure}[h!]
    \centering
    \hspace{-20pt}
    \begin{minipage}{.3\textwidth}
        \centering
        \includegraphics[width=5.5cm]{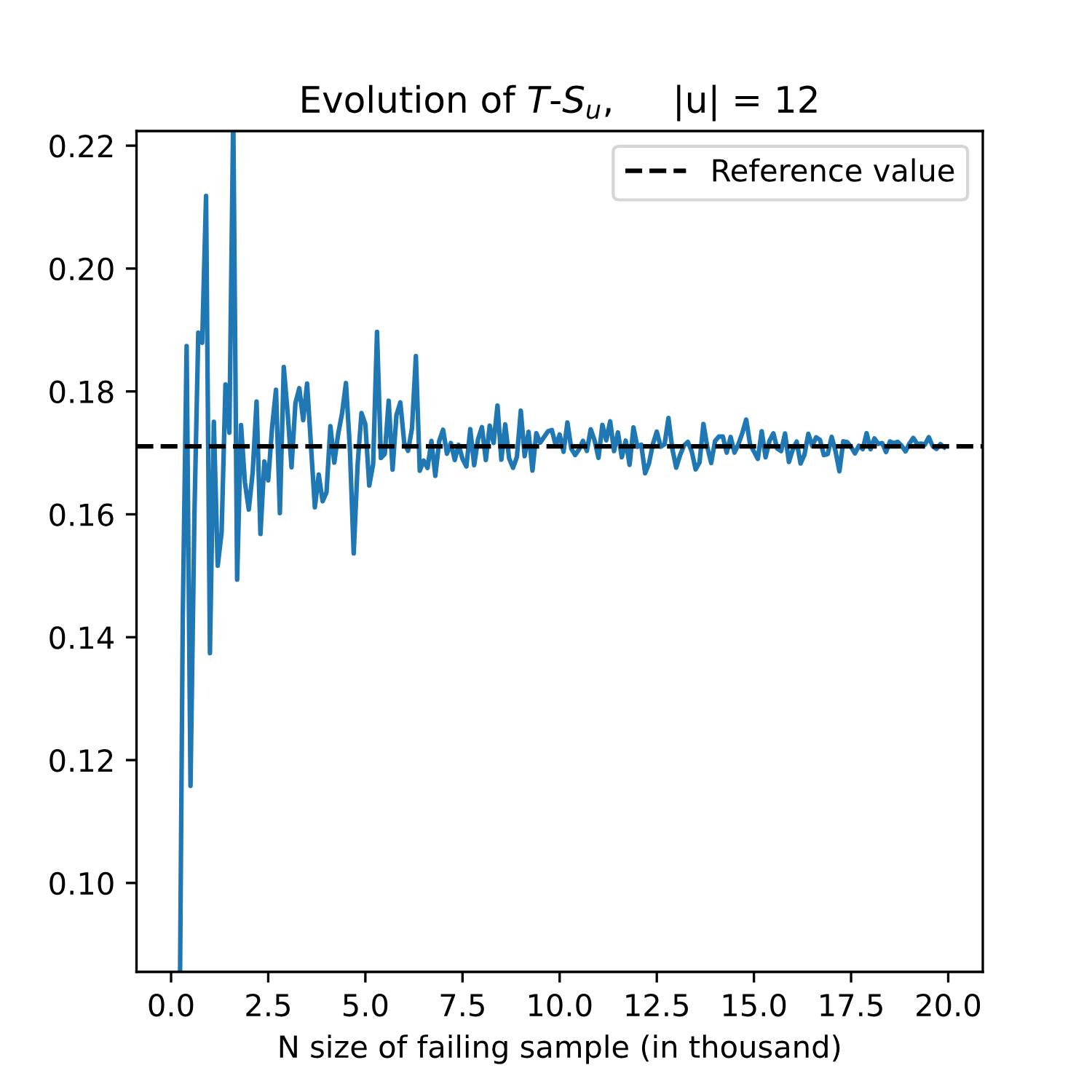}
        {\footnotesize $(a)$ \\ $|u| = 12$}
    \end{minipage}
    \hspace{10pt}
    \begin{minipage}{0.3\textwidth}
        \centering
        \includegraphics[width=5.5cm]{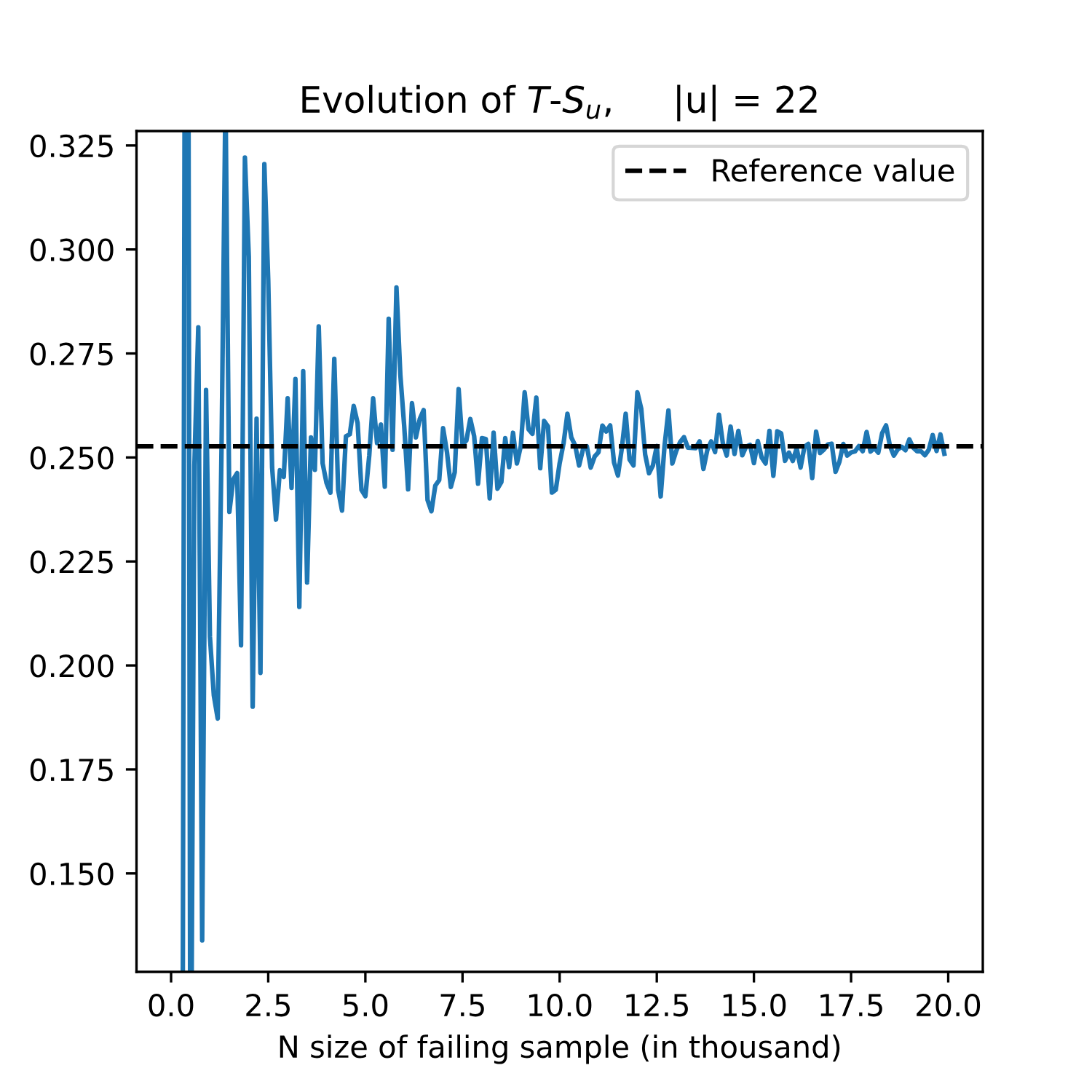}
        {\footnotesize $(b)$ \\ $|u| = 22$}
    \end{minipage}
    \hspace{10pt}
    \begin{minipage}{0.3\textwidth}
        \centering
       \includegraphics[width=5.5cm]{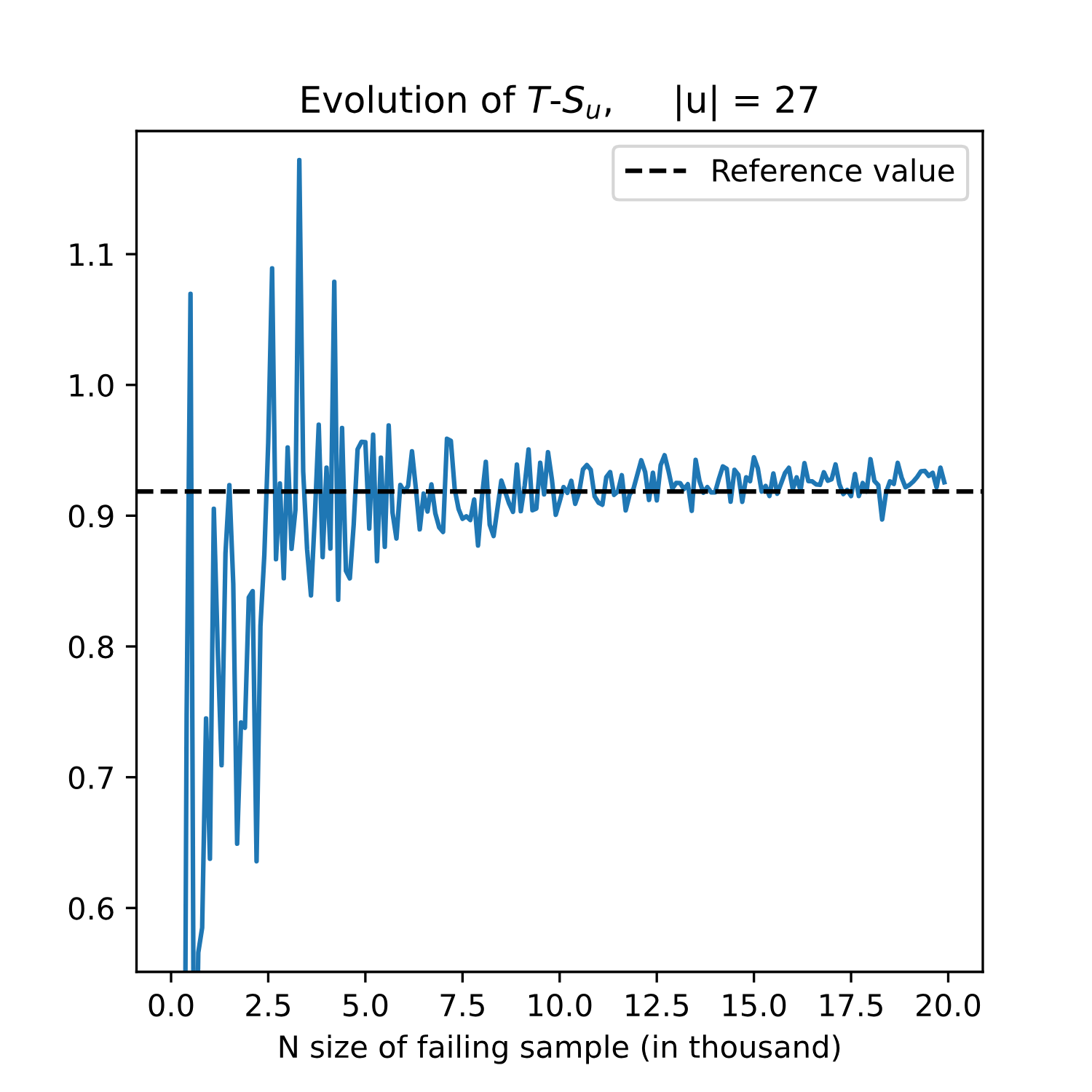}
        {\footnotesize $(c)$ \\ $|u| = 27$}
    \end{minipage}
    \caption{Evolution of the estimate of the closed target Sobol indices when the failing sample size $N$ varies from $100$ to $20 \ 000$.}
    \label{fig:fig3_size_sample}
\end{figure}
\end{center}

We can make two important remarks regarding the Figure \ref{fig:fig1_2estim}: in both cases (the blue curve and the orange curve), the estimate roughly stabilizes once $2 \ 000$ points have been used to estimate the quantity under study (given that $10 \ 000$ points have been used for the other quantity), and increasing the number of points allows the estimate to converge towards the reference value for the three considered Sobol indices. With regard to the study of the parameter $\alpha$ in Figure \ref{fig:fig2_alpha}, the estimate appears to be the most stable and accurate around $\alpha = 0.5$, which corresponds to an equipartition of the failing sample for estimating the density $f_{\mathbf X_u | F_t}$ and the expectation $\text{E}_u$. Based on these results, we set $\alpha = 1/2$ and we proceed to a third analysis which consists in studying the impact of the failing sample size $N$ on the estimation of $\text{T-S}_u$. We illustrate in Figure \ref{fig:fig3_size_sample} the estimation of the three considered closed target Sobol indices for $N \in \Iintv{100, 20 \ 000}\}$. For the three considered target Sobol indices, the estimation roughly stabilizes after $N \approx 5 \ 000$ or $N \approx 6 \ 000$.

\begin{center}
    \begin{figure}[h!]
    \centering
    \hspace{-20pt}
    \begin{minipage}{.3\textwidth}
        \centering
        \includegraphics[width=5.5cm]{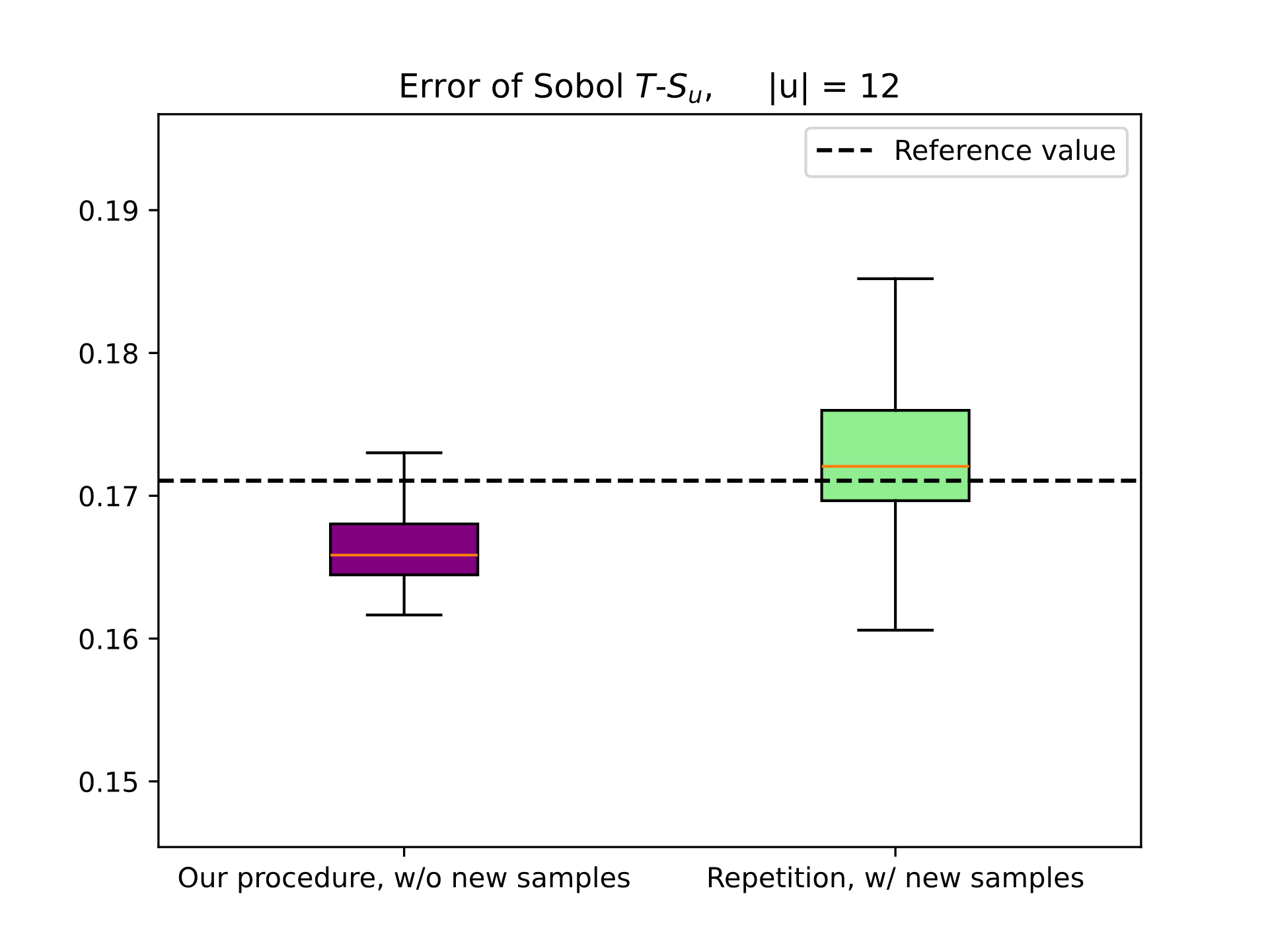}
        {\footnotesize $(a)$ \\ $|u| = 12$}
    \end{minipage}
    \hspace{10pt}
    \begin{minipage}{0.3\textwidth}
        \centering
        \includegraphics[width=5.5cm]{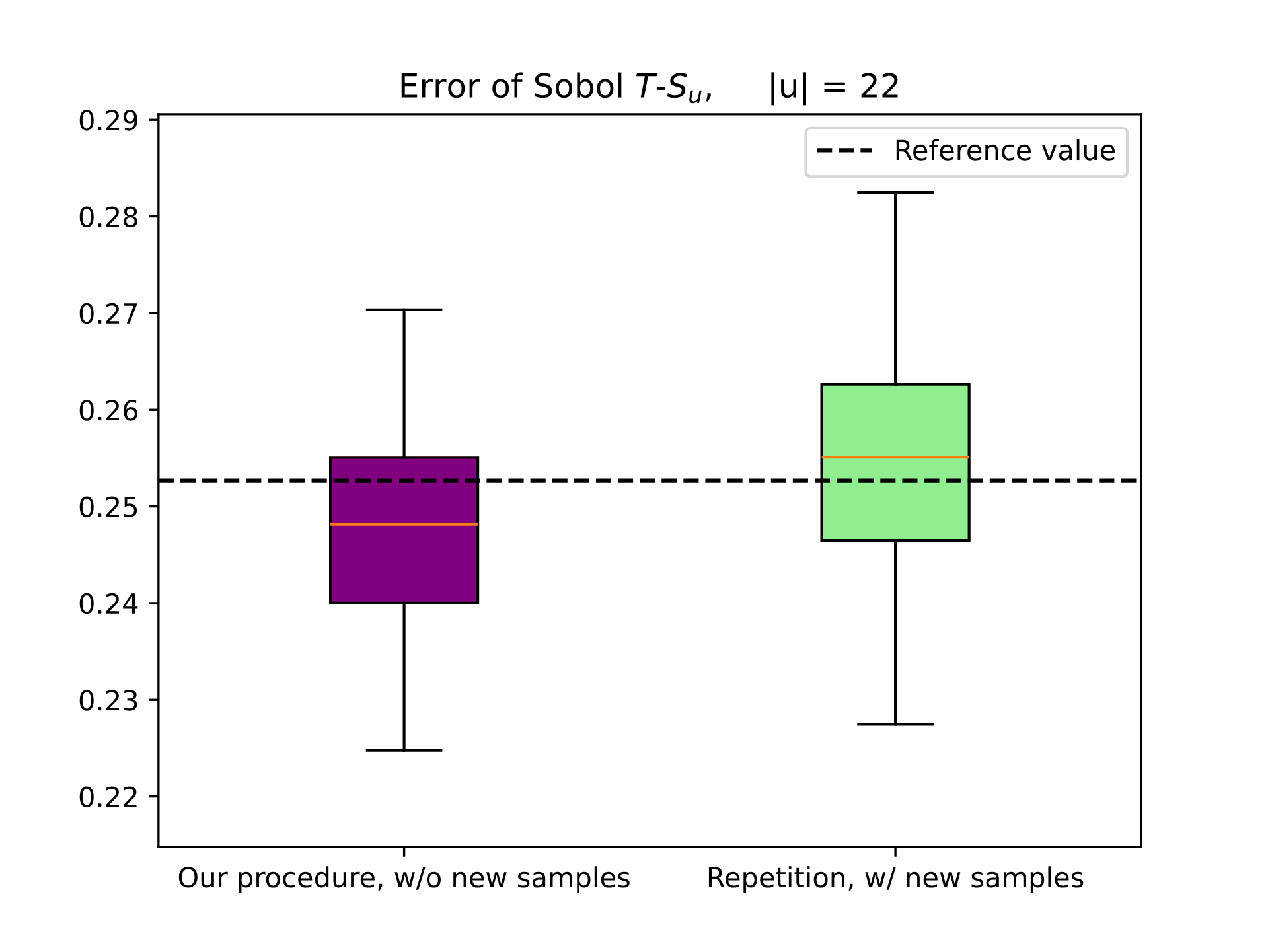}
        {\footnotesize $(b)$ \\ $|u| = 22$}
    \end{minipage}
    \hspace{10pt}
    \begin{minipage}{0.3\textwidth}
        \centering
       \includegraphics[width=5.5cm]{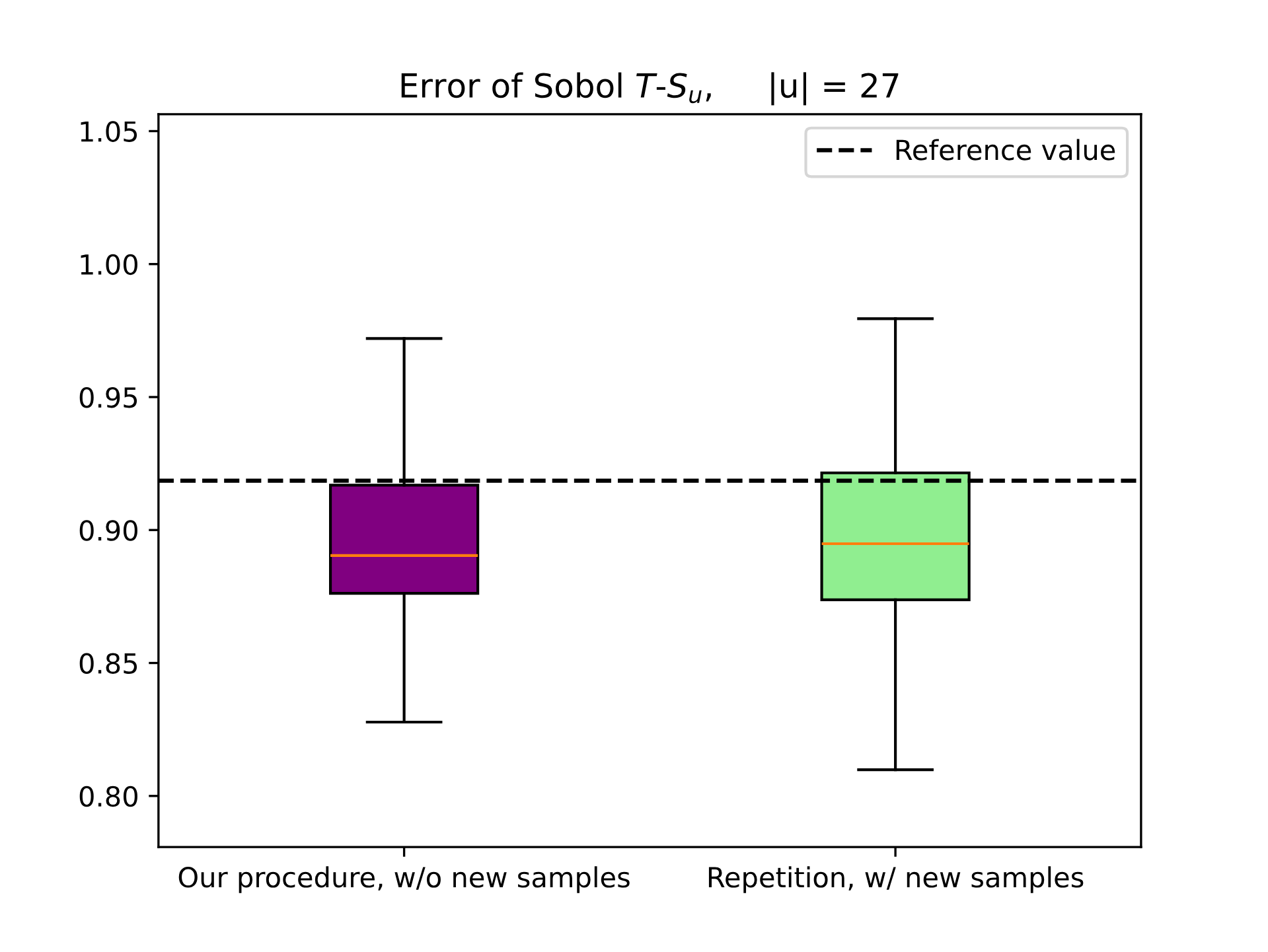}
        {\footnotesize $(c)$ \\ $|u| = 27$}
    \end{minipage}
    \caption{Estimation of closed target Sobol indices with its error quantified by our procedure (using a unique sample of  $5 \ 000$ failing points) \textit{vs.} with $100$ repetitions of the estimation (using each time a new failing sample of size $5 \ 000$).}
    \label{fig:fig4_error_sobol}
\end{figure}
\end{center}

Next, in Figure \ref{fig:fig4_error_sobol}, we illustrate our uncertainty estimation procedure for the target Sobol indices, as developed in Section \ref{sec:err_sobol}, where we compare the reference value, the error obtained using our procedure, and the error obtained by repeating the estimation with a new initial sample each time. The aim here is to verify whether our procedure accurately captures the uncertainty in the estimate of $\text{T-S}_u$. We now consider a failing sample of size $5 \ 000$, and given the result in Figure \ref{fig:fig3_size_sample}, this should create a visible uncertainty on the estimation of the target Sobol indices. We set $J=100$ for Algorithm \ref{algo:uq_sobol} to obtain the uncertainty with our procedure, and we carry out $100$ repetitions with new failing samples of size $5 \ 000$ to obtain the uncertainty with repetition of the estimation. As before, we use a GMM with $5$ components to estimate the conditional densities. We observe that our procedure can recover quite accurately the uncertainty of the target Sobol indices, although the uncertainty can be a bit underestimated as it does not fully take into account the randomness on the sample since it is based on a resampling procedure.

\begin{center}
    \begin{figure}[h!]
        \centering
        \includegraphics[width=\linewidth]{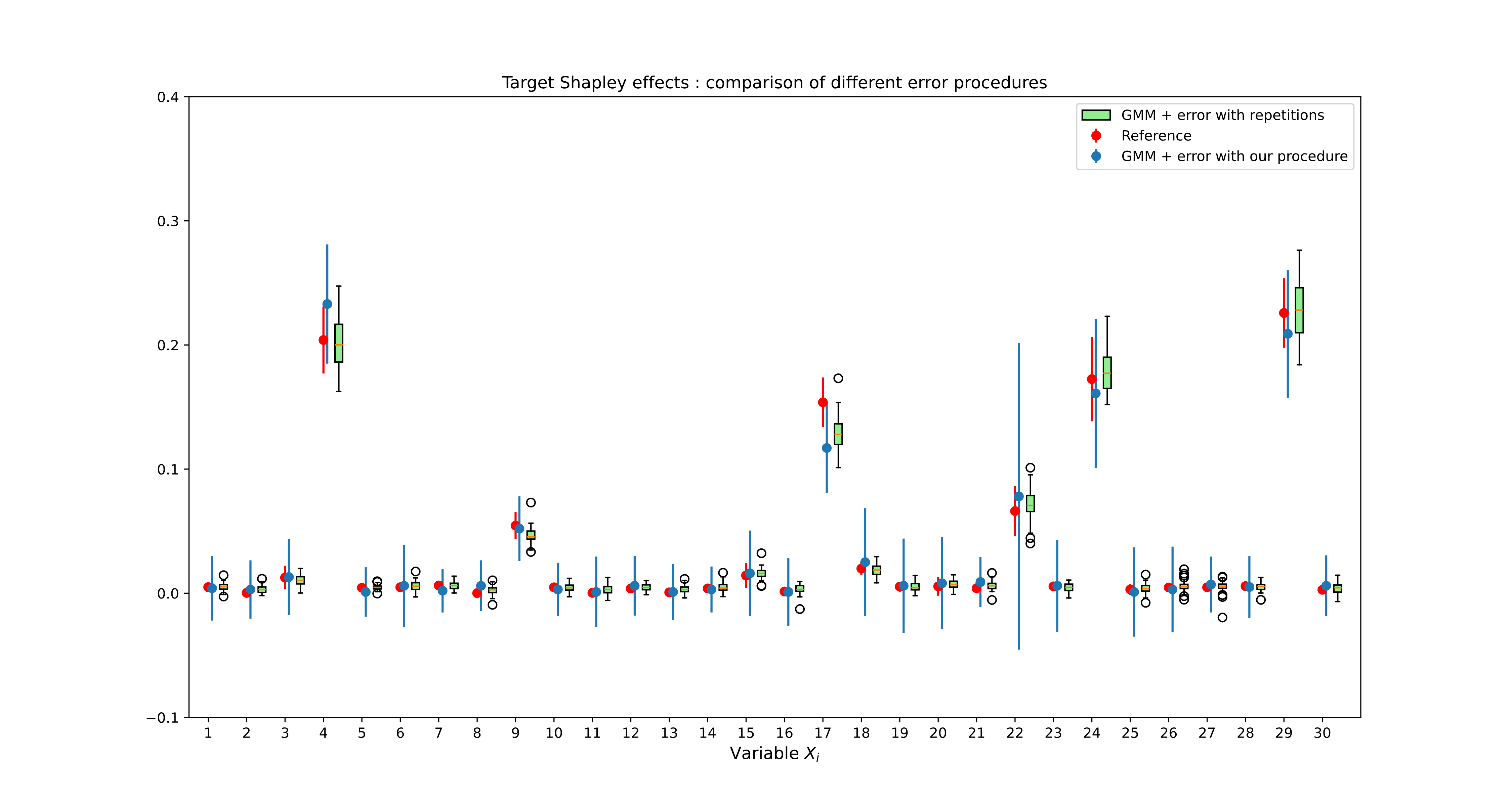}
    \caption{Estimation of the target Shapley effects with their error for different procedures: $95\%$ confidence intervals with reference values for the closed target Sobol indices (in red), $95\%$ confidence intervals with estimated values for the closed target Sobol indices (in blue), distribution of 100 repetitions of the estimation (in green).}
    \label{fig:fig5_error_shapley}
\end{figure}
\end{center}

We now illustrate in Figure \ref{fig:fig5_error_shapley} the estimation of $\text{T-Sh}_i$, $1 \leq i \leq d$, along with their error. First of all, we set $\delta = 0.05$ to consider $95\%$ confidence intervals. To begin with, we uniformly sample $300$ permutations to calculate the reference values. Since we calculate the reference target Sobol indices, there is a small approximation error due to the sampling of these $300$ permutations for the calculation of the reference values of the target Shapley effects. These values and their errors are shown in red in Figure \ref{fig:fig5_error_shapley}, and can be interpreted to mean that the reference values lie within the red intervals with probability $95\%$. Next, we uniformly sample $M=100$ permutations and we estimate the corresponding closed target Sobol indices by applying Algorithm \ref{algo:uq_marg_contri} using only the failing sample of size $N=5,000$ and setting $J=10$, $L=10$ and we use a GMM with 5 components as before. We can recover estimates $\widehat{r}_{(i)}$, $1 \leq i \leq d$, and we build the associated $95\%$ confidence intervals (in blue in Figure \ref{fig:fig5_error_shapley}). Finally, we also compare in Figure \ref{fig:fig5_error_shapley} $100$ repetitions of estimation of the target Shapley effects, each repetition using a different sample of $5 \ 000$ failing points and a different set of $100$ permutations. We can make several observations. First, the standard repeated-sampling estimate performs well, with one small exception for $\text{T-S}_{17}$, which seems more difficult to estimate, although the order of magnitude is estimated correctly. Second, we note that our procedure tends to yield an estimate of the target Shapley effects that is offset from the reference value, which can be explained by the $M=100$ permutations sampled. We also note that the error tends to be overestimated compared to the repeated-sampling estimate (especially ofr $X_{22}$), which can be explained by the fact that the estimated range $\widehat{r}_{(i)}$ can be large if one of the underlying estimated Sobol indices is relatively poorly estimated, thereby providing a poorly estimated marginal contribution (possibly too small or too large). Since the error bound is a probabilistic bound, the interval is large to ensure the threshold of $1 - \delta$.

\begin{center}
    \begin{figure}[h!]
        \centering
        \includegraphics[width=\linewidth]{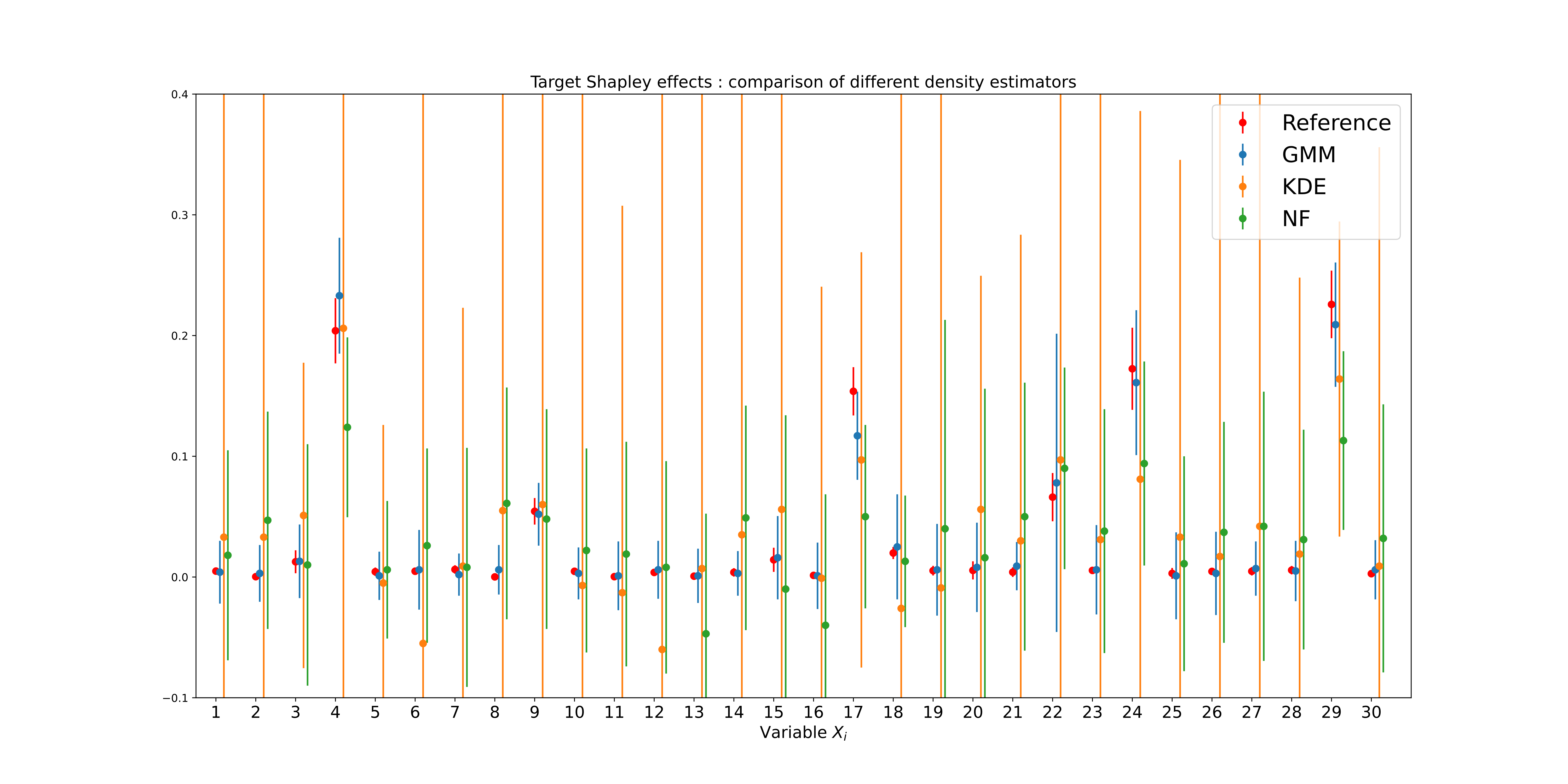}
    \caption{Estimation of the target Shapley effects with their error for different estimators : GMM 5-components, KDE and Normalizing Flows.}
    \label{fig:fig6_comp_shapley}
\end{figure}
\end{center}

Finally, we compare in Figure \ref{fig:fig6_comp_shapley} the estimation of $\text{T-Sh}_i$, $1 \leq i \leq d$ for several density estimators: KDE, GMM and normalizing flows (NF). The choice of normalizing flows is motivated by the fact that these density estimators perform well in high dimensions \cite{coccaro2024robusttestnf} and are fairly flexible, but they require a sufficiently important number of data points and are quite computationally intensive, which is exacerbated the estimation is computed for several subgroups $u \in \mathcal P(d)$. We present them here only for the sake of comparison. The same sample of $100$ permutations is used for all three estimations and we provide the computation times for each of the aforementioned estimators in Table \ref{tab:tab2_computing_time}. We observe that NF have difficulty to estimate the large values of target Shapley effects and exhibit large confidence intervals, KDE provides huge confidence interval and GMM is globally the most precise estimator. We observe that our procedure to quantify the error is sensible since one inaccurate estimation may provide a large range $r_{(i)}$, as shown for the estimation with GMM for the variable $X_{22}$.

\begin{table}[h!]
    \centering
    \begin{tabular}{c|c|c|c|}
        Estimator & GMM & KDE & NF \\
        Computing time (hh:mm:ss) & 1:25:45 & 00:56:04 & 26:15:20 \\
    \end{tabular}
    \caption{Comparison of computation times for the three considered density estimators.}
    \label{tab:tab2_computing_time}
\end{table}

\subsection{Fire-spread case}

As a second example, we consider a problem frequently studied in the literature on Shapley effects \cite{song2016shapley, demangechryst2023shaprosa}: the forest fire spread model. This model was initially introduced by Rothermel \cite{rothermel1972firespread}, and was subsequently modified \cite{albini1976wildfire, catchpole1991modellingmoisture, salvador2001gsafire, song2016shapley}. We refer to \cite{song2016shapley} and \cite{demangechryst2023shaprosa} for additional details regarding the system of equations governing the model. This model has 10 input variables: $\mathbf X = (\delta, \sigma, h, \rho_p, m_l, m_d, S_T, U, \tan \varphi, P)$ and two output variables, but the one of interest to us corresponds to the fire spread rate at a specific point, given in $cm \cdot s^{-1}$, and denoted by $R$. The marginal distributions of the inputs and their interpretation are given in Table \ref{tab:tab_distrib_fs}.

For a numerical perspective, truncated distributions are used in order to consider the following modifications:

\begin{itemize}
    \item For all the input variables, the negative values are removed.
    \item For the variables $S_T$ and $P$, the values greater than $1$ are removed.
    \item For the variable $m_d$, the values lower than $3 / 0.6$ are removed.
\end{itemize}

\begin{table}[]
    \centering
    \small
    \begin{tabular}{llll}
        \hline
        n° & Input variable & Symbol (unit) & Distribution \\
        \hline
        1  & Fuel depth  &  $\delta$ ($cm$)   &  $\log \mathcal N (2.19, 0.517)$ \\
        2  & Fuel particle area-to-volume ratio  &  $\sigma$ ($cm^{-1}$)  &  $\log \mathcal N (3.31, 0.294)$ \\
        3  & Fuel particle low heat content  &  $h$ ($Kcal \cdot kg^{-1}$)   &  $\log \mathcal N (8.48, 0.063)$ \\
        4  & Oven-dry particle density  &  $\rho_p$ ($D \cdot W \cdot g \cdot cm^{-3}$)  &  $\log \mathcal N (-0.592, 0.219)$ \\
        5  & Moisture content of the live fuel  &  $m_l$ ($H_2OgD \cdot W \cdot g^{-1}$)  &  $\mathcal N (1.18, 0.377)$ \\
        6  & Moisture content of the dead fuel  &  $m_d$ ($H_2OgD \cdot W \cdot g^{-1}$)  &  $\mathcal N (0.19, 0.047)$ \\
        7  & Fuel particle total mineral content  &  $S_T$  ($MIN \cdot gD \cdot W \cdot g^{-1}$)  &  $\mathcal N (0.049, 0.011)$ \\
        8  & Wind speed at midflame height  &  $U$ ($km \cdot h^{-1}$)  &  $\log \mathcal N (2.9489, 0.5569)$ \\
        9  & Slope  &  $\tan \varphi$   &  $\mathcal N (0.38, 0.186)$ \\
        10 & Dead fuel loading to total fuel loading  &  $P$  &  $\log \mathcal N (-2.19, 0.64)$\\
        \hline
    \end{tabular}
    \caption{Description of input variables for the fire-spread model. $\mathcal N(\mu, \sigma)$ denotes a 1D Gaussian distribution with mean $\mu \in \mathbb R$ and standard deviation $\sigma \in \mathbb R_+$ and $a \log \mathcal N(\mu, \sigma)$ denotes the distribution of $a \exp (I)$ with $I \sim \mathcal N(\mu, \sigma)$.}
    \label{tab:tab_distrib_fs}
\end{table}

In addition to these marginal distributions, observations in \cite{clark2008firespreadsa} suggest a negative correlation between the variables $m_d$ and $U$, and we assume that this correlation is given by the Pearson correlation coefficient $corr(m_d, U) = -0.8$. The dependence structure of $(m_d, U)$ is modeled by a Gaussian copula. Furthermore, the critical failure threshold is set at $t = 60 \, cm \cdot s^{-1}$, which means that we seek to identify the variables explaining extreme dispersion of forest fires, according to the model. An estimation of the failure probability is given by $p_t \approx 10^{-4}$, which is obtained using a Monte Carlo simulation with a sample size of $10^7$, and the reference values of the target Shapley effects are obtained using a double Monte Carlo estimator from the literature that uses a much larger sample budget than the other estimators compared here. We refer to \cite{demangechryst2023shaprosa} for additional details. For the estimation of the target Shapley effects with our estimator, we rely on a failing sample of size $N=5400$ obtained from a Monte Carlo simulation.

\begin{center}
    \begin{figure}[h!]
        \centering
        \includegraphics[width=\linewidth]{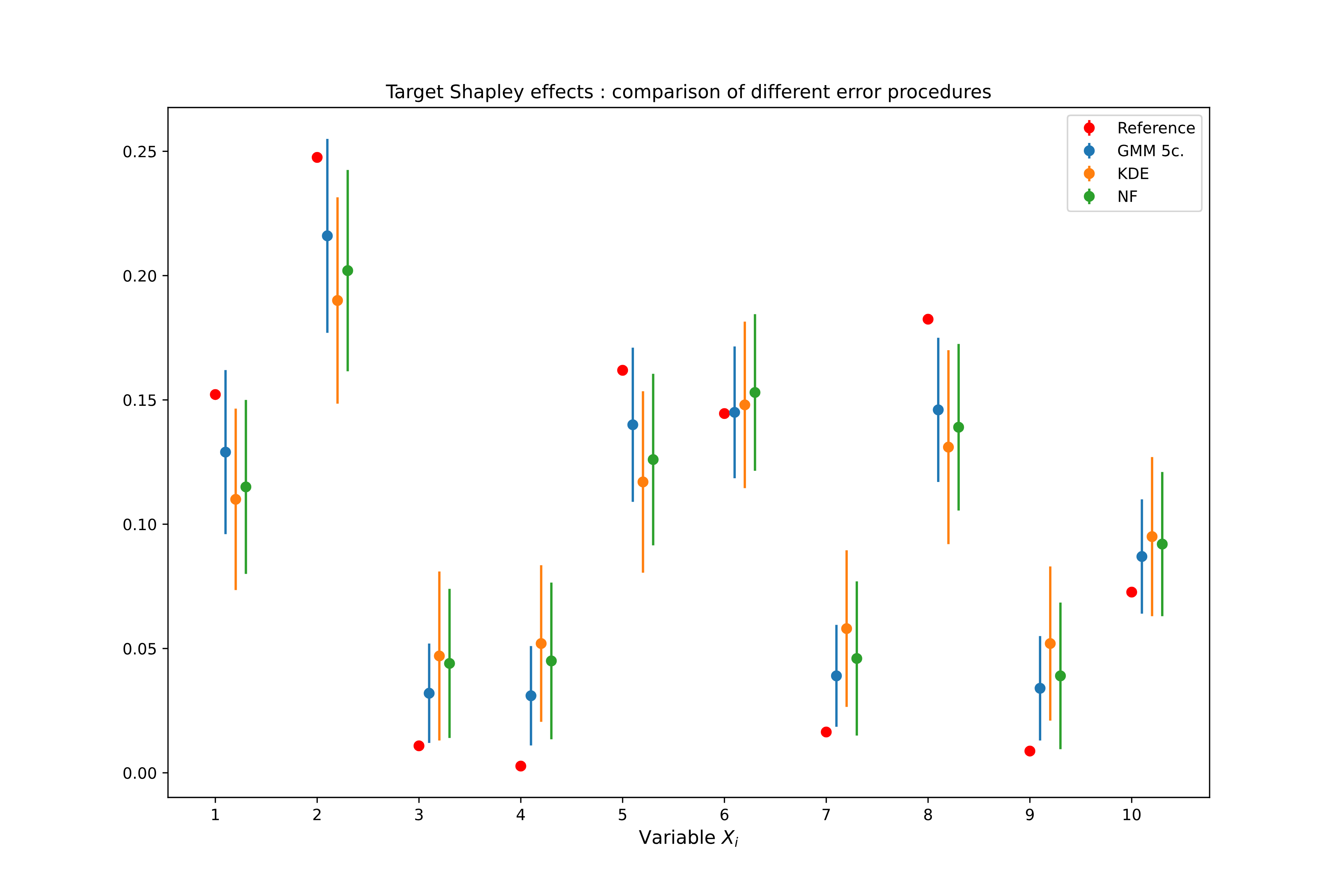}
        \caption{\textbf{Fire-spread case results.} Reference values \textit{vs.} estimation of target Shapley effects with GMM (+ our error estimation), with KDE (+ our error estimation), and with NF (+ our error estimation).}
    \label{fig:fig7_fs_comp}
    \end{figure}
\end{center}

As before, we estimate the target Shapley effects and the estimation error using NF, GMM with $5$ components, and KDE. We use a sample of $M=200$ permutations, we set $J=10$, $L=10$, and $\delta=0.05$. The same sample of $200$ permutations is used for all three estimations, as before. The results are shown in Figure \ref{fig:fig7_fs_comp} and the computing times are provided in Table \ref{tab:tab3_computing_time_fs}. It seems that all the three density estimators agree on the final estimates for the $10$ target Shapley effects, but it also seems that they are limited and cannot reach the reference values. This may be explained by the fact that the three density estimators make a similar mistake in their estimation due to too complex input densities, or by the fact that the considered sample of $M$ permutations is not appropriate and that a better sample should be chosen.

\begin{table}[h!]
    \centering
    \begin{tabular}{c|c|c|c|}
        Estimator & GMM & KDE & NF \\
        Computation time (hh:mm:ss) & 1:13:15 & 00:53:12 & 12:41:59 \\
    \end{tabular}
    \caption{Comparison of computation times for the three considered density estimators.}
    \label{tab:tab3_computing_time_fs}
\end{table}

\subsection{Application to Adaptive Importance Sampling}

In this section, we provide a brief overview of adaptive importance sampling (AIS) and show how our method can be used to monitor the impact of input variables on changes in probability. The idea is to quantify, at each step of the algorithm, the role of each variable in the occurrence of the threshold exceedance.

Adaptive importance sampling \cite{beh2025} consists in iteratively update an intermediate threshold and the auxiliary density sampling distribution in order to get closer to the failure region, and thus to the threshold $t$. The idea is to choose a family of densities $\mathcal Q$ and to successively generate a sample $\mathbf X^{(n)}_ {k+1} \sim q^{(k+1)}$, where the density $q^{(k+1)} \in \mathcal Q$ was itself estimated using the previous sample $\mathbf X^{(n)}_{k} \sim q^{(k)}$, with $q^{(k)} \in \mathcal Q$.

To be more precise, to update $q^{(k+1)}$ we set an initial parameter $\rho \in [0,1]$ such that at each iteration, if $N$ points $\mathbf X_{k}^{(n)} \sim q^{(k)}$ are generated, we keep the $\lfloor \rho N \rfloor$ points that exceed the $(1-\rho)$-quantile $\gamma_{k} = \min(t, \Phi(\mathbf X^{(\lceil (1-\rho) N \rceil)}_{k}))$. Then we update the auxiliary density $q^{(k+1)}$ based on these $\lfloor \rho N \rfloor$ points with an adaptation rule $\mathcal A$ such as cross entropy for parametric families \cite{papaioannou2019, wang2016}. Therefore, at each step $k$, we have a known density $q^{(k)}$ according to which we sample, and a region corresponding to a threshold exceedance $\Gamma_{k} = \{x \in \mathcal X : \Phi(x) > \gamma_{k}\}$. To simplify the notation, we set $\Gamma = \Gamma_k$ for all $k$, and denote by $q^{(k)}_{\mathbf X | \Gamma}$ the density $q^{(k)}_{\mathbf X}$ conditional on the realization $\Gamma$. Note also that, by definition, the event $\Gamma$ has probability $\rho$ for any $k$ (except at the last iteration). Furthermore, the variables $X_1, \ldots, X_d$ are not necessarily independent at step $k$. We provide in \ref{appendix:ais_algo} a summary of the algorithm of Adaptive Importance Sampling for the estimation of failure probability.

Based on these definitions, we can consider our estimation scheme for target Shapley effects, and we can quantify the impact of the variables $X_1, \ldots, X_d$ on the realization of $\Gamma$ at each iteration. The objective, then, is to monitor at each step how the input variables affect the convergence toward the optimal density $q_\star \propto \mathbf 1_{f(x) > t}(x) f(x)$. Our formulation using densities allows us to directly address this monitoring problem, since that each step $k$ the density $q^{(k)}_{\mathbf X}$ is available (corresponding to the base density at step $k$) and the points $\mathbf X_k^{(n)} \sim q^{(k)}_{\mathbf X | \Gamma}$ that realize $\Gamma$ are also available (the $\lfloor \rho N \rfloor$ elite points), which is suitable to apply our methodology. In addition, estimating the target Shapley effects at each step requires no additional calls to the model $\Phi$. Hence for no additional cost we can obtain a better understanding of the role played by each variable in the AIS procedure. However, care must be taken with the interpretation, since the target Shapley effects obtained only allow to quantify the impact of a variable when moving from step $k$ to step $k+1$.

Hence, we define the closed target Sobol index associated with the group $u \in \mathcal P(d)$ at step $k$ by

\begin{equation}\label{eq:target_sobol_ais}
    \text{T-S}_{u}^{(k)} = \frac{\rho}{1 - \rho} \left( \mathbb E_{q^{(k)}_{\mathbf X_u | \Gamma}} \left[ \frac{q^{(k)}_{\mathbf X_u | \Gamma}(\mathbf X_u)}{q^{(k)}_{\mathbf X_u}(\mathbf X_u)}\right] - 1 \right), \qquad \mathbf X_u \sim q^{(k)}_{\mathbf X_u | \Gamma}.
\end{equation}

To simplify notations, we have omitted the notation with the superscript $\text{c}$. We can thus apply our methodology at each iteration of the AIS to compute the target Shapley effects associated with exceeding $\gamma_k$ for all $k$.

We now illustrate the previous discussion on two use cases. First we consider a toy case in dimension $d=20$, with the following nonlinear function

\begin{equation}
    \Phi(\mathbf X) = X_1 + X_2^2 + 5 X_3 X_4 + 0.1 X_6 + X_7 + X_9^3 + 2\cos(\pi X_{10}) + X_{11} + 5 X_{13} + 0.04 (X_{16} - X_{20})^4 + 20
\end{equation}

\begin{center}
    \begin{figure}[h!]
        \centering
        \includegraphics[width=8cm]{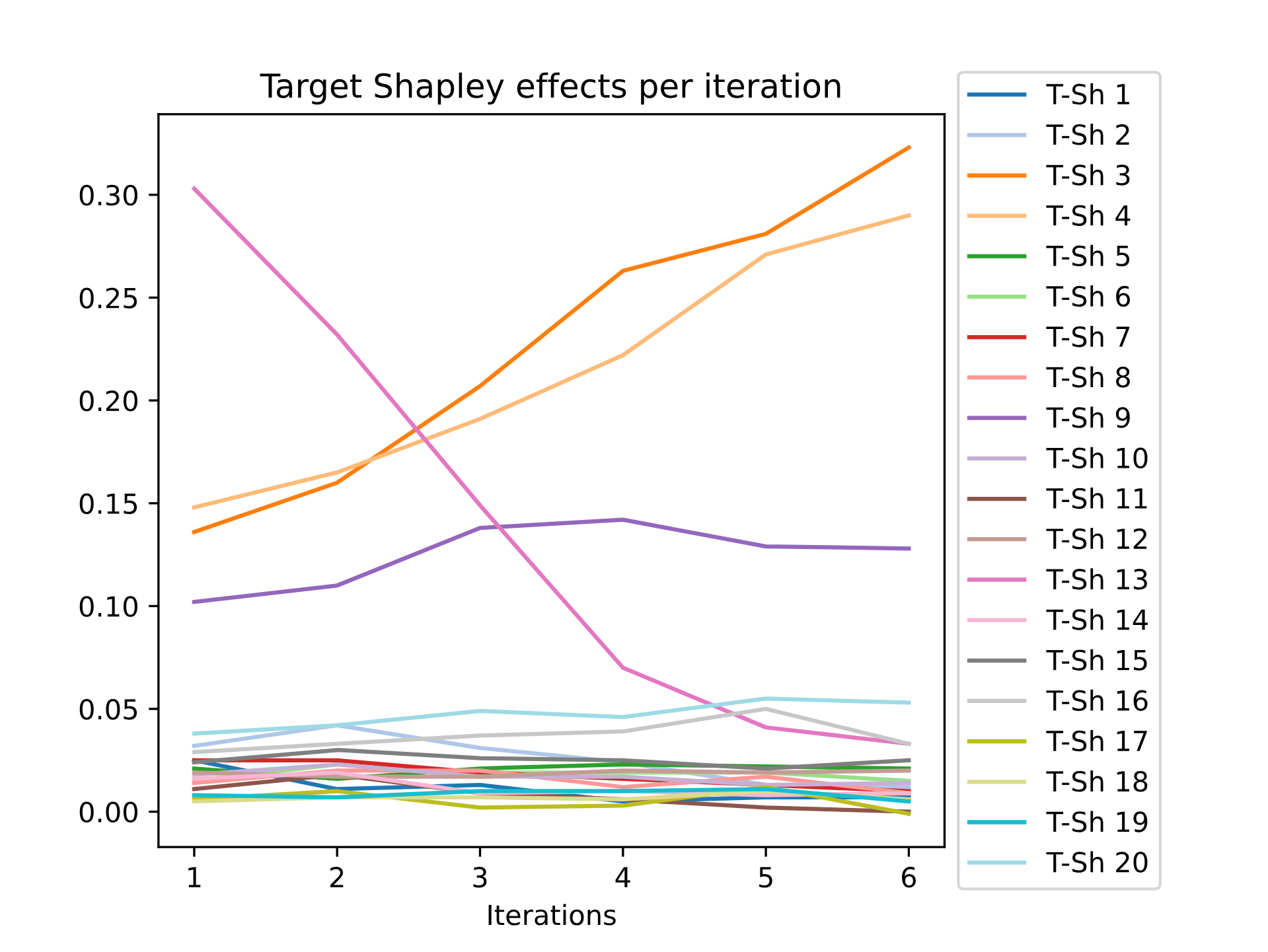}
        \caption{Evolution of the target Shapley effects through the AIS iterations for the nonlinear toy case.}
    \label{fig:fig8_ais_toycase}
    \end{figure}
\end{center}

We suppose that the input variables are mutually independent and normally distributed and we have omitted the variables $X_5$, $X_8$, $X_{12}$, $X_{14}$, $X_{15}$, $X_{17}$, $X_{18}$ and $X_{19}$ in the model. We set the threshold to $t=100$ so that the failure probability is $p_t \approx 10^{-5}$. We consider Gaussian auxiliary densities, we sample $10 \ 000$ points per iteration and we set $\rho = 0.3$ so that $3 \ 000$ points are kept at each iteration $k$ to estimate $\theta_k$. This also means that $3 \ 000$ points are used at each iteration to estimate the corresponding target Shapley effects. Moreover, we sample $M=100$ permutations and set $J=5$, $L=5$ in Algorithm \ref{algo:uq_marg_contri}. We estimate the target Shapley effects at each iteration with $5$-components GMM, using the points exceeding the threshold for the correspond iteration. The results are displayed in Figure \ref{fig:fig8_ais_toycase} and the computing time is provided in Table \ref{tab:tab4_computing_time_ais1}. The variable $X_{13}$ has a major role for the first iterations, which quickly decreases, and the variable finally has no impact in the occurrence of the threshold exceeding in the last iterations. On the contrary, the variables $X_{3}$ and $X_{4}$ have impact right from the first iterations and this impact conjointly increases, which can be explained by the fact that $X_3$ and $X_4$ are related in the function $\Phi$. Finally, the variable $X_9$ has roughly the same impact on the threshold exceeding throughout the entire estimation scheme.

\begin{table}[h!]
    \centering
    \begin{tabular}{c|c}
        Computation time (hh:mm:ss) & 1:35:04 \\
    \end{tabular}
    \caption{Computation time for the estimation of target Shapley effects for all iterations, for the nonlinear toy case.}
    \label{tab:tab4_computing_time_ais1}
\end{table}

\begin{center}
    \begin{figure}[h!]
    \centering
    \hspace{-20pt}
    \begin{minipage}{.45\textwidth}
        \centering
        \includegraphics[width=7cm]{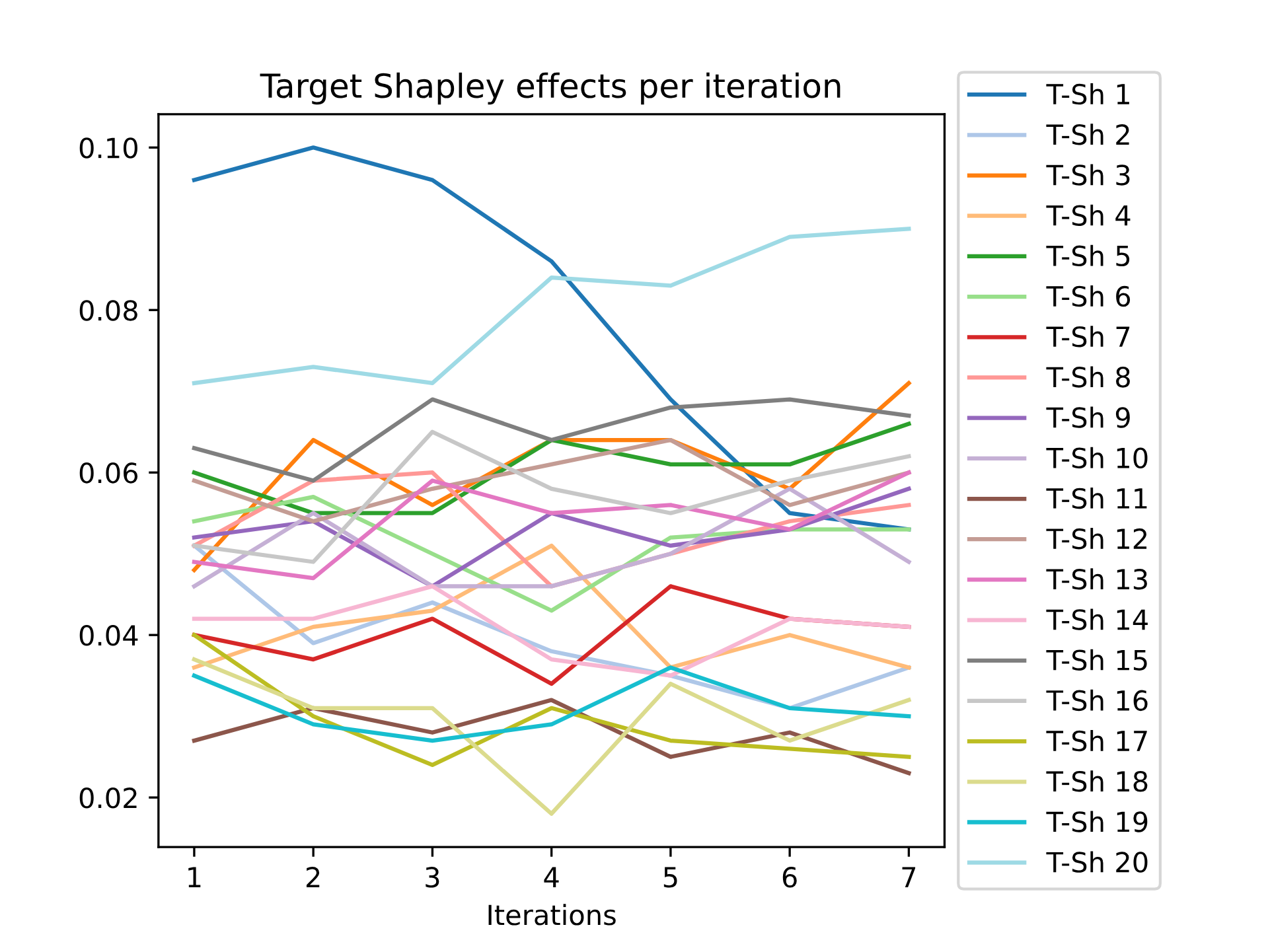}
        {\footnotesize First computation.}
    \end{minipage}
    \hspace{20pt}
    \begin{minipage}{0.45\textwidth}
        \centering
        \includegraphics[width=7cm]{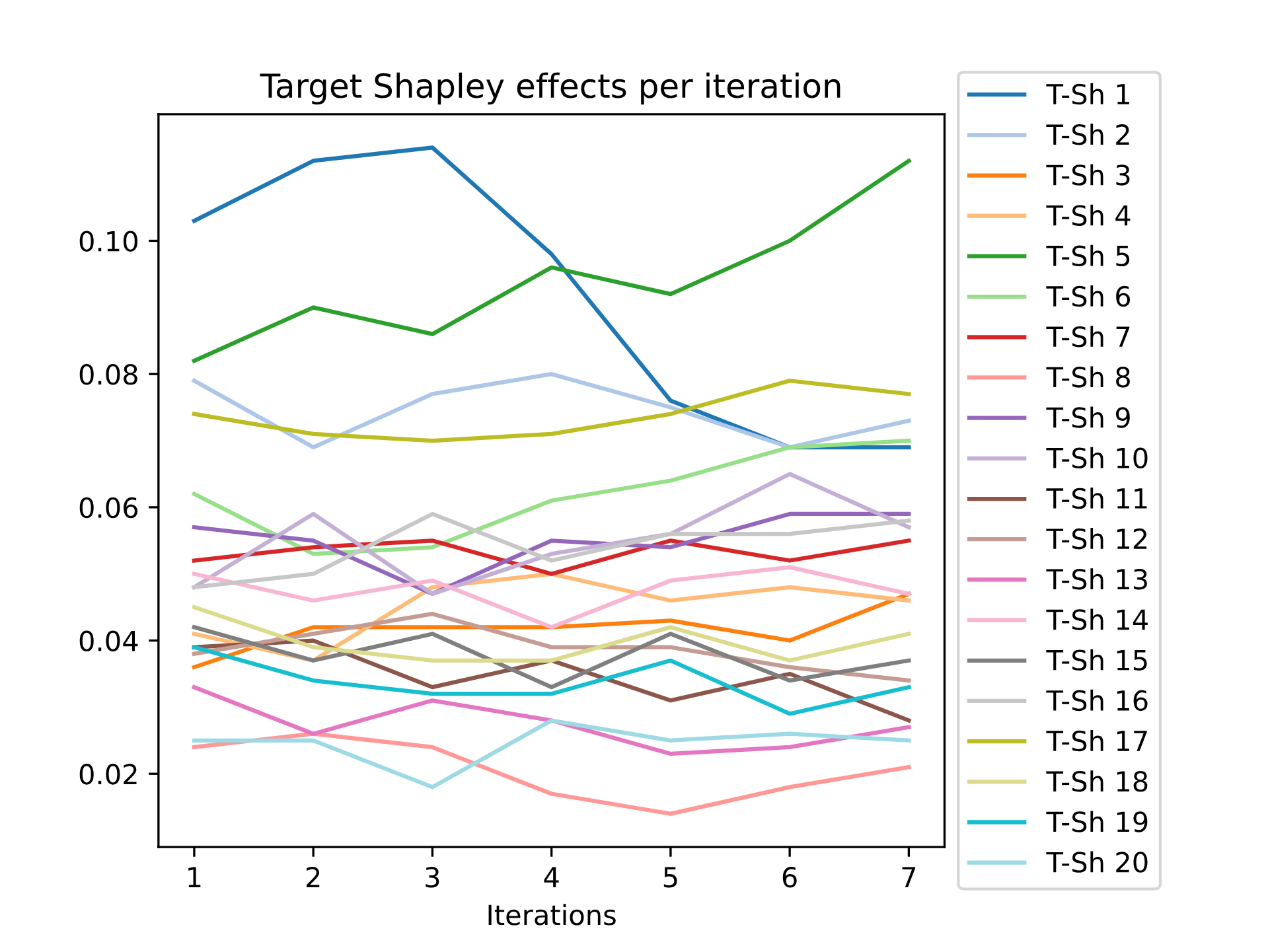}
        {\footnotesize Second computation.}
    \end{minipage}
    \caption{Evolution of the target Shapley effects for the exceedance threshold, computed at each iteration, for the portfolio use-case.}
    \label{fig:fig9_ais_portfolio}
\end{figure}
\end{center}

We now consider a second use case, given by the following application in finance \cite{bassamboo2008, chan2012, elmasri2024}. We consider a $(n+2)$-dimensional portfolio loss function defined by

\begin{equation}
    \Phi(\mathbf X) = \sum_{j=3}^{n+2} \mathbf 1_{\left\{\Lambda(X_1, X_2, X_j) \geq 0.5 \sqrt{n}\right\}} - bn,
\end{equation}

where $b$ is a constant and $\Lambda$ is the function defined by

\begin{equation}
    \Lambda(X_1, X_2, X_j) = \left( q X_1 + 3 (1-q^2)^{1/2} X_j \right) \left( F_\Gamma^{-1} (F_{\mathcal N}(X_2)) \right)^{-1/2},
\end{equation}

with $F_\Gamma$ and $F_{\mathcal N}$ respectively the cumulative distribution functions of $\Gamma(6,6)$ and $\mathcal N(0,1)$. The uncertainty on the $(n+2)$-dimensional input random vector $\mathbf X$ is modeled by the $(n+2)$-dimensional standard Gaussian distribution, and the failure threshold is $t=0$. We choose the constant $b=0.7$ in order to obtain $p \approx 10^{-4}$. We set the dimension to $n+2 = 20$, we sample $M=100$ permutations and set $J=5$, $L=5$ in Algorithm \ref{algo:uq_marg_contri}, we use a $5$-components GMM as before to estimate the densities, we set $\rho = 0.2$ and we sample $10 \ 000$ points per iterations. The results are displayed in Figure \ref{fig:fig9_ais_portfolio} and the computing time are provided in Table \ref{tab:tab5_computing_time_ais2}. The variables $X_3, X_4, \ldots, X_{20}$ should have globally the same impact, while $X_1$ and $X_2$ should have their own behavior. We perform the estimation procedure two times in order to identify the respective behavior of each variable from its computational noise. We observe in Figure \ref{fig:fig9_ais_portfolio} that $X_1$ plays an important role in the early iterations, which rapidly diminishes, and that $X_2$ plays the same role throughout the iterations. The values between the two computations are not exactly the same (\textit{i.e.} $\text{T-Sh}_1$), but the variations are the same. Finally, the other variables $X_3, X_4, \ldots, X_{20}$ do not exhibit any clear or predominant behavior in causing the thresholds $\gamma_k$ to be exceeded, as expected.

\begin{table}[h!]
    \centering
    \begin{tabular}{c|c|c}
                            & $1^{st}$ computation & $2^{nd}$ computation \\
        Computation time (hh:mm:ss) & 2:14:38 & 2:06:21
    \end{tabular}
    \caption{Computation time for the estimation of target Shapley effects for all iterations for the portfolio use case.}
    \label{tab:tab5_computing_time_ais2}
\end{table}

\section{Conclusion} \label{sec:conclusion}

In this article, we tackled the challenge of estimating the reliability-oriented Shapley effects for high-dimensional models when the input variables are dependent. These sensitivity indices are particularly useful as they allow to quantify the influence of the model inputs on the occurrence of the failure, even when they are correlated. Hence, allowing to estimate these indices in higher-dimensional settings enables their use for a wider range of real-world applications.

We propose a methodology based on a $N$-sample of failing points. Once such a sample has been recovered, along with an estimation of the failure probability, one can apply our procedure to obtain estimates of the target Shapley effects as well as an estimation of the error using the same $N$-sample. To do so, we have relied on a new writing of the closed target Sobol indices involving marginal densities conditionally to the failure event. Once these densities have been estimated with the appropriate tools, we approximate the target Shapley effects using permutations sampling. In addition, we have proposed a procedure to quantify the error made on the estimation of the target Shapley effects, with the same $N$-sample provided by the reliability analysis. This is an important contribution for the practitioner as it avoids to repeat the estimation scheme like several other methods, hence avoiding additional calls to the model. Finally, we have illustrated our methodology on several applications, in particular to the monitoring of adaptive importance sampling scheme for tracking the effects of the input variables on threshold exceedance at each step. This last application does not require any additional call to the model once the AIS has been achieved.

Further perspectives can consist in improving the permutations sampling. Indeed the computational burden is mainly due to the number of permutations to sample, and our error estimation procedure is heavily constrained by the drawn sample of $M$ permutations. It might be worthwhile to integrate existing methods such as Quasi Monte Carlo \cite{mitchell2022perm} or importance sampling \cite{benard2022shaff} to better select this sample of permutations. The error estimate would then need to be adapted to account for these sampling methods. Finally, an interesting perspective would be to study our estimate of the target Shapley effects from a theoretical perspective by considering an appropriate density estimator which theoretical analysis of the error is tractable, and propagating its error. Although one strength of our method is the use of any density estimator, we nevertheless lack theoretical guarantees regarding the resulting estimate and its true error.

\section*{Acknowledgements}

Our work has benefitted from the AI Interdisciplinary Institute ANITI. ANITI is funded by the France 2030 program under the Grant agreement n°ANR-23-IACL-0002.

\newpage

\appendix

\section{Integral representations of \eqref{eq:target_first_sobol} and \eqref{eq:target_total_sobol}} \label{appendix:integralexp}

For the sake of precision and clarity, we provide the integral representations of \eqref{eq:target_first_sobol} and \eqref{eq:target_total_sobol}.

First, for the first-order target Sobol indices, we have

\begin{equation}
    \begin{split}
    \mathbb V_{f_{X_i}} \left(\frac{f_{X_i | F_t}(X_i)}{f_{X_i}(X_i)}\right)
        &= \mathbb E_{f_{X_i}} \left[\left(\frac{f_{X_i | F_t}(X_i)}{f_{X_i}(X_i)}\right)^2\right] - \mathbb E_{f_{X_i}} \left[\frac{f_{X_i | F_t}(X_i)}{f_{X_i}(X_i)}\right]^2 \\
        &= \int_{\mathcal X_i} \frac{f_{X_i | F_t}(x_i)^2}{f_{X_i}(x_i)^2} f_{X_i}(x_i) dx_i - \left[\int_{\mathcal X_i} \frac{f_{X_i | F_t}(x_i)}{f_{X_i}(x_i)} f_{X_i}(x_i) dx_i\right]^2.
    \end{split}
\end{equation}

Second, for the total-order target Sobol indices, we have

\begin{equation}
    \begin{split}
    \mathbb V_{f_{\mathbf X_{-i}}} \left(\frac{f_{\mathbf X_{-i} | F_t}(\mathbf X_{-i})}{f_{\mathbf X_{-i}}(\mathbf X_{-i})}\right)
        &= \mathbb E_{f_{\mathbf X_{-i}}} \left[\left(\frac{f_{\mathbf X_{-i} | F_t}(\mathbf X_{-i})}{f_{\mathbf X_{-i}}(\mathbf X_{-i})}\right)^2\right] - \mathbb E_{f_{\mathbf X_{-i}}} \left[\frac{f_{\mathbf X_{-i} | F_t}(\mathbf X_{-i})}{f_{\mathbf X_{-i}}(\mathbf X_{-i})}\right]^2 \\
        &= \int_{\mathcal X_{-i}} \frac{f_{\mathbf X_{-i} | F_t}(\mathbf x_{-i})^2}{f_{\mathbf X_{-i}}(\mathbf x_{-i})^2} f_{\mathbf X_{-i}}(\mathbf x_{-i}) d\mathbf x_{-i} - \left[\int_{\mathcal X_{-i}} \frac{f_{\mathbf X_{-i} | F_t}(\mathbf x_{-i})}{f_{\mathbf X_{-i}}(\mathbf x_{-i})} f_{\mathbf X_{-i}}(\mathbf x_{-i}) d\mathbf x_{-i}\right]^2,
    \end{split}
\end{equation}

where $d\mathbf x_{-i}$ means that we integrate with respect to all variables except $x_i$.

\section{Proofs of \eqref{eq:rewriting_var} and \eqref{eq:rewriting_exp}} \label{appendix:proofs}

This section details the proofs of the rewritings in \eqref{eq:rewriting_var} and \eqref{eq:rewriting_exp}.

For the first one, the numerator can be written as

\begin{equation}
    \begin{split}
        & \mathbb V_{f_{\mathbf X_u}} \left(\mathbb E [\mathbf 1_{F_t}(\mathbf X) |  \mathbf X_u]\right) \\
        &= \mathbb E_{f_{\mathbf X_u}} \left[ \left( \mathbb E [\mathbf 1_{F_t}(\mathbf X) |  \mathbf X_u] - \mathbb E_{f_{\mathbf X_u}}\left[ \mathbb E [\mathbf 1_{F_t}(\mathbf X) |  \mathbf X_u] \right] \right)^2 \right] \\
        &= \mathbb E_{f_{\mathbf X_u}} \left[ \left( \int_{\mathcal X_{-u}} \mathbf{1}_{F_t}(\mathbf X_u, \mathbf x_{-u}) f_{\mathbf X_{-u} | \mathbf X_u} (\mathbf x_{-u}) d\mathbf x_{-u} - \int_{\mathcal X_u} \left( \int_{\mathcal X_{-u}} \mathbf{1}_{F_t}(\mathbf X_u, \mathbf x_{-u}) f_{\mathbf X_{-u} | \mathbf X_u} (\mathbf x_{-u}) d\mathbf x_{-u} \right) f_{\mathbf X_u} (\mathbf x_{u}) d\mathbf x_{u} \right)^2 \right] \\
        &= \mathbb E_{f_{\mathbf X_u}} \left[ \left( \mathbb P(\Phi(\mathbf X) > t \, | \, \mathbf X_u) - \mathbb P(\Phi(\mathbf X) > t) \right)^2 \right],
    \end{split}
\end{equation}

where $d\mathbf x_{u}$ means that we integrate with respect to all variables in $u$, and $d\mathbf x_{-u}$ means that we integrate with respect to all variables not in $u$.

Relying on Bayes' theorem, we obtain

\begin{equation}
    \mathbb P(\Phi(\mathbf X) > t \, | \, \mathbf X_u) = \mathbb P(\Phi(\mathbf X) > t) \frac{f_{\mathbf X_u | \Phi(\mathbf X) > t}(\mathbf X_u)}{f_{\mathbf X_u}(\mathbf X_u)} = p_t \frac{f_{\mathbf X_u | F_t}(\mathbf X_u)}{f_{\mathbf X_u}(\mathbf X_u)},
\end{equation}

hence providing

\begin{equation}
    \begin{split}
        \mathbb E_{f_{\mathbf X_u}} & \left[ \left( \mathbb P(\Phi(\mathbf X) > t \, | \, \mathbf X_u) - \mathbb P(\Phi(\mathbf X) > t) \right)^2 \right] \\
        &= \mathbb E_{f_{\mathbf X_u}} \left[ \left( p_t\frac{f_{\mathbf X_u | F_t}(\mathbf X_u)}{f_{\mathbf X_u}(\mathbf X_u)}  - p_t\right)^2 \right] \\
        &= p_t^2 \mathbb E_{f_{\mathbf X_u}} \left[ \left( \frac{f_{\mathbf X_u | F_t}(\mathbf X_u)}{f_{\mathbf X_u}(\mathbf X_u)} - 1 \right)^2 \right] \\
        &= p_t^2 \left( \mathbb V_{f_{\mathbf X_u}} \left( \frac{f_{\mathbf X_u | F_t}(\mathbf X_u)}{f_{\mathbf X_u}(\mathbf X_u)} - 1 \right) + \left( \mathbb E_{f_{\mathbf X_u}} \left[ \frac{f_{\mathbf X_u | F_t}(\mathbf X_u)}{f_{\mathbf X_u}(\mathbf X_u)} - 1 \right] \right)^2 \right)\\
        &= p_t^2 \left( \mathbb V_{f_{\mathbf X_u}} \left(\frac{f_{\mathbf X_u | F_t}(\mathbf X_u)}{f_{\mathbf X_u}(\mathbf X_u)} \right) + \mathbb E_{f_{\mathbf X_u}} \left[\frac{f_{\mathbf X_u | F_t}(\mathbf X_u)}{f_{\mathbf X_u}(\mathbf X_u)} \right]^2 - 2\mathbb E_{f_{\mathbf X_u}} \left[\frac{f_{\mathbf X_u | F_t}(\mathbf X_u)}{f_{\mathbf X_u}(\mathbf X_u)} \right] + 1 \right).
    \end{split}
\end{equation}

Since $\mathbb E_{f_{\mathbf X_u}} \left[\frac{f_{\mathbf X_u | F_t}(\mathbf X_u)}{f_{\mathbf X_u}(\mathbf X_u)} \right] = \int_{\mathcal X_u} \frac{f_{\mathbf X_u | F_t}(\mathbf x_u)}{f_{\mathbf X_u}(\mathbf x_u)} f_{\mathbf X_u}(\mathbf x_u) d\mathbf x_u =  \int_{\mathcal X_u} f_{\mathbf X_u | F_t}(\mathbf x_u) d\mathbf x_u = 1$, we obtain

\begin{equation}
        \mathbb V_{f_{\mathbf X_u}} (\mathbb E [\mathbf 1_{F_t}(\mathbf X) |  \mathbf X_u]) = p_t^2 \left( \mathbb V_{f_{\mathbf X_u}} \left(\frac{f_{\mathbf X_u | F_t}(\mathbf X_u)}{f_{\mathbf X_u}(\mathbf X_u)} \right)\right).
\end{equation}

Finally, since $\mathbf 1_{F_t}(\mathbf X) \sim \mathcal B(p_t)$, we have $\mathbb V (\mathbf 1_{F_t}(\mathbf X)) = p_t(1-p_t)$ and thus

\begin{equation}
    \frac{\mathbb V_{f_{\mathbf X_u}} (\mathbb E [\mathbf 1_{F_t}(\mathbf X) |  \mathbf X_u])}{\mathbb V(\mathbf 1_{F_t}(\mathbf X))} = \frac{p_t}{1-p_t} \left( \mathbb V_{f_{\mathbf X_u}} \left(\frac{f_{\mathbf X_u | F_t}(\mathbf X_u)}{f_{\mathbf X_u}(\mathbf X_u)} \right)\right).
\end{equation}

This concludes the proof of \eqref{eq:rewriting_var}.

To obtain the expectation in \eqref{eq:rewriting_exp}, we have

\begin{equation} 
    \begin{split}
        \mathbb V_{f_{\mathbf X_u}} \left( \frac{f_{\mathbf X_u | F_t}(\mathbf X_u)}{f_{\mathbf X_u}(\mathbf X_u)} \right)
        &= \mathbb E_{f_{\mathbf X_u}}\left[\left( \frac{f_{\mathbf X_u | F_t}(\mathbf X_u)}{f_{\mathbf X_u}(\mathbf X_u)} \right)^2\right]  -\mathbb E_{f_{\mathbf X_u}}\left[ \frac{f_{\mathbf X_u | F_t}(\mathbf X_u)}{f_{\mathbf X_u}(\mathbf X_u)} \right]^2\\
        &= \int_{\mathcal X_u} \left( \frac{f_{\mathbf X_u | F_t}(\mathbf x_u)}{f_{\mathbf X_u}(\mathbf x_u)} \right)^2 f_{\mathbf X_u}(\mathbf x_u) d\mathbf x_u - 1^2\\
        &= \int_{\mathcal X_u} \mathbf 1_{F_t}(\mathbf x_u, \mathbf x_{-u}) \frac{f_{\mathbf X_u | F_t}(\mathbf x_u)}{f_{\mathbf X_u}(\mathbf x_u)} f_{\mathbf X_u | F_t}(\mathbf x_u) d\mathbf x_u - 1 \\
        &= \mathbb E_{\mathbf X_u | F_t}\left[\frac{f_{\mathbf X_u | F_t}(\mathbf X_u)}{f_{\mathbf X_u}(\mathbf X_u)} \right] - 1.
    \end{split}
\end{equation}

This concludes the proof of \eqref{eq:rewriting_exp}.

\vspace{30pt}

\section{Adaptive Importance Sampling Algorithm for Reliability Analysis}\label{appendix:ais_algo}

In this appendix we recall the AIS algorithm for reliability analysis, when we consider Gaussian auxiliary densities. Thus the density $q_{\theta}$ is a Gaussian density of parameter $\theta = (\mu, \Sigma)$. $\mathcal A$ denotes a generic adaptation rule. $\mathcal A$ takes $\mathbf X_k^{(n)}$ with their weights $w_n$ for the elite points $n \in I_k$ and $\theta_k$, and outputs $\theta_{k+1} = (\mu_{k+1}, \Sigma_{k+1})$.

\begin{algorithm}[H] \label{algo:ais_fail}
\SetKwData{Left}{left}\SetKwData{This}{this}\SetKwData{Up}{up}
\SetKwFunction{Union}{Union}\SetKwFunction{FindCompress}{FindCompress}
\SetKwInOut{Input}{input}\SetKwInOut{Output}{output}
\Input{nominal density $f_{\mathbf X}$, function $\Phi$, threshold $t$, parametric family $\{q_\theta\}_{\theta \in \Theta}$,
      adaptation rule $\mathcal A$, sample size $N$, proportion $\rho \in (0,1)$, initial parameter $\theta_0 \in \Theta$,
      final sample size $N_{\mathrm{IS}}$, maximum number of iterations $k_{\max}$}
\Output{$\widehat p_t^{IS}$}
\BlankLine
    $k \leftarrow 0$\;
\Repeat{$\gamma_{k-1} = t$ \textbf{or} $k = k_{\max}$}{
  Generate $\mathbf X_k^{(1)}, \dots, \mathbf X_k^{(N)} \overset{\text{iid}}{\sim} q_{\theta_k}$\;
  Evaluate $\Phi(\mathbf X_k^{(n)})$ and reindex so that
  $\Phi(\mathbf X_k^{(1)}) \leq \dots \leq \Phi(\mathbf X_k^{(N)})$\;
  $\gamma_k \leftarrow \min\big(t,\ \Phi(\mathbf X_k^{(\lceil (1-\rho) N \rceil)})\big)$
  \tcp*{empirical $(1-\rho)$-quantile}
  $I_k \leftarrow \{ n : \Phi(\mathbf X_k^{(n)}) \geq \gamma_k \}$
  \tcp*{elite points in $\Gamma_k$}
  \For{$n \in I_k$}{
    $w_n \leftarrow f_{\mathbf X}(\mathbf X_k^{(n)}) \,/\, q_{\theta_k}(\mathbf X_k^{(n)})$\;
  }
  $\theta_{k+1} \leftarrow \mathcal A\big(\{(\mathbf X_k^{(n)}, w_n)\}_{n \in I_k},\ \theta_k\big)$
  \tcp*{adaptation step}
  $k \leftarrow k+1$\;}
  $K \leftarrow k$\;
  Generate $\mathbf X^{(1)}, \dots, \mathbf X^{(N_{\mathrm{IS}})} \overset{\text{iid}}{\sim} q_{\theta_K}$\;
  $\displaystyle \widehat p_t^{IS} \leftarrow \frac{1}{N_{\mathrm{IS}}} \sum_{n=1}^{N_{\mathrm{IS}}} \mathbf 1_{F_t}(\mathbf X^{(n)}) \, \frac{f_{\mathbf X}(\mathbf X^{(n)})}{q_{\theta_K}(\mathbf X^{(n)})}$\;
 \caption{Adaptive importance sampling for estimating $p_t = \mathbb P(\Phi(\mathbf X) > t)$}
\end{algorithm}

\newpage
\bibliographystyle{elsarticle-harv} 
\bibliography{thebibliography}

\end{document}